\documentclass[12pt]{article}

\usepackage{natbib}
\bibpunct{(}{)}{;}{a}{}{,}
\usepackage{setspace}

\usepackage{amsfonts,amsmath, bbold}
\usepackage{graphicx}

\newcommand{\bbeta}{\boldsymbol \beta}
\newcommand{\bgamma}{\boldsymbol \gamma}

\newcommand{\ump}[1]{$\! \! ~\ref{#1}$}

\newcommand{\umpp}[1]{($\! \! ~\ref{#1}$)}

\newcommand{\qand}{\quad \mbox{and} \quad}

\newcommand{\qfor}{\quad \mbox{for} \quad}

\newcommand{\bbR}{{\mathbb R}}

\newcommand{\ep}{\epsilon}
\newcommand{\bx}{{\bf x}}

\newcommand{\by}{{\bf y}}
\newcommand{\bz}{{\bf z}}

\newcommand{\dspl}{\displaystyle}
\newcommand{\sol}[2]{#1/#2}

\newcommand{\qw}[1]{}

\newcommand{\donotusepdffigures}[1]{}
\newcommand{\mr}[1]{}

\newcommand{\bv}{{\bf v}}

\newcommand{\bzero}{{\bf 0}}
\newcommand{\ith}{{\mbox{\scriptsize{th}}}}

\begin{document} 

\newcommand{\blind}{0}

\newcommand{\tit}{\bf The Plateau Problem in the
    Heteroskedastic Probit Model}
    
\if0\blind

{\title{\tit\thanks{We thank Michael Alvarez and John Brehm for kindly providing us with
their data set. For comments and suggestions, we thank Chris Achen, Neal Beck, John Brehm, Andrew
Gelman, Gary King, Andrew Martin, Walter Mebane, Kevin Quinn, Jasjeet Sekhon, Jim Stimson, Marco Steenbergen, Greg Wawro, Richard Williams, Fred Boehmke, and the Statistics Working Group at Columbia University. We thank Andrew Gelman for making this collaboration possible by introducing three of the authors.}}


\author{Eric Freeman\thanks{Institute for Research on Labor and Employment, University of California at Berkeley}
\and Luke Keele\thanks{Department of Political Science, Penn State University, Corresponding Author}
\and David Park\thanks{Applied Statistics Center, Columbia University}
\and Julia Salzman\thanks{Department of Biochemistry and Department of Statistics, Stanford University}
\and Brendan Weickert\thanks{Palantir Technologies}
}

\date{\today}

\maketitle
}\fi

\if1\blind
\title{\bf \tit}
\maketitle
\fi 

\thispagestyle{empty}

\begin{abstract}
In parameter determination for the heteroskedastic probit model, both in
simulated data and in actual data collected by \citet{Alvarez:1995b}, we observe a failure of traditional local search methods to converge consistently to a single parameter vector, in contrast to
the typical situation for the regular probit model.  We identify
features of the heteroskedastic probit log likelihood function that we argue tend to lead to this failure, and suggest ways to amend the local search methods to remedy the problem.
\end{abstract}

\clearpage

\doublespacing

A common difficulty encountered in the estimation of statistical models
is that of unequal variances or heteroskedasticity. In the context of ordinary
least squares, heteroskedasticity does not bias parameter
estimates; rather, it either inflates or underestimates the standard
errors. Heteroskedasticity, however, is more problematic in discrete
choice models such as logit or probit and their ordered and
multinomial variants.  If we have nonconstant variances in the error
term of a discrete choice model, not only are the standard errors
incorrect, but the parameters are also biased and inconsistent
\citep{Yatchew:1985}.

\citet{Alvarez:1995b} generalize a basic model for heteroskedasticity developed by \citet{Harvey:1976}.
They call these heterogenous choice models, which include heteroskedastic probit and heteroskedastic ordered probit models to correct for unequal variances with discrete outcomes. They also used heteroskedasticity as a means of exploring heterogeneity in choice situations.  These heteroskedastic probit models have been frequently used to explore heterogenous choices and behaviors \citep{Alvarez:1997, Alvarez:1998c, Alvarez:2002, Busch:1999, Gabel:1998, Lee:2002, Krutz:2005}. Routines for these models have become standard in statistical software such as \textsf{Stata}, \textsf{Limdep}, \textsf{SAS}, \textsf{Eviews} and \textsf{Shazaam}.  

We use simulations to explore the optimization difficulties that can
arise during the estimation of these models. In our simulations, the
models are correct. Thus, the anomalies that we find cannot be explained
by specification error. We argue that estimation difficulties are due
to the functional form of these models. In this paper, we observe the following dichotomy: the usual algorithms for optimization of regular probit log likelihood often fail with the heteroskedastic probit model. It is likely that the problems with these model are under-reported in the literature, since the errors that are encountered may give rise to the ``file-drawer'' problem where difficulties with these models go unreported as these papers are sitting in investigators' file drawers \citep{Iyengar:1988}.
The point of this paper is, first, to give evidence of that dichotomy, second, to attempt to identify those features of the heteroskedastic likelihood function that lead to the failure, and third to suggest ways a local search might be adapted to those features.

To this end, in section \ref{sec:models} of this paper, we give a brief exposition of
the regular and heteroskedastic probit models.  In section \ref{sec:performance}, we
investigate the performance of ``out-of-the-box'' optimization techniques with the heteroskedastic probit model, first through simulated data, and second through a data set collected by \citet{Alvarez:1995b}.  In section \ref{sec:intuition}, we delve more deeply into our simulated data to attempt to gain some intuition about the graph of the likelihood function, and thus about the failure of local search methods, in the heteroskedastic case in general.  And in section \ref{sec:suggestion}, we suggest a modification to tailor optimization methods to the heteroskedastic probit model, and offer additional recommendations.

\section{A Probit Model with Heteroskedasticity}
\label{sec:models}

We first define the model, beginning with the standard probit model. Fix a constant
$k_1-$dimensional vector $\bbeta$, and assume that, given an observed value $\bx_i$
of a $k_1-$dimensional vector of random variables and an observed value
$\ep_i$ of a random variable distributed normally with mean $0$
and variance $\dspl{\sigma^2}$, a latent random
variable $\dspl{y^{\ast}}$ takes the value\footnote{To be completely precise, we note that
  we think of vectors as column vectors.}
\[ y_i^{\ast} = \bx_i' \bbeta + \ep_i. 
\]
This value is not seen by the researcher. A researcher observes only the
value
\[ y_i = \left\{ \begin{array}{cc}
    1 & \mbox{if  } \ y_i^{\ast} \geq 0 \\
    0 & \mbox{if  } \ y_i^{\ast} < 0. \\
    \end{array} \right.
\]
One then readily calculates that 
\[ \mbox{Pr}(y_i=1 \mid \bx_i) = \Phi \left( \frac{\bx_i' \bbeta}{\sigma} \right),
\]
where $\Phi$ is the cumulative distribution function for the standard
normal distribution. Only the parameter $\dspl{\frac{1}{\sigma}\bbeta}$
is identified, so typically one assumes
\[ \sigma = 1
\]
for purposes of identification.  

In the heteroskedastic probit model, one allows $\sigma$ to take values
other than $1$, by estimating a model for its value. This may arise in
many situations, for example, in a model of personal choice where levels of 
information vary across individuals. A convenient form for modeling $\sigma$ is
\[ \sigma = \exp( \bz'\bgamma ),
\]
where $\bz$ is a $k_2-$dimensional random vector and
$\bgamma$ is a $k_2-$dimensional parameter vector. Thus the model can be
written as 
\[ \mbox{Pr}(y_i=1) = \Phi \left( \frac{\bx_i' \bbeta}{\exp( \bz'\bgamma )} \right).
\]
A natural way to think of this model, as above, is as consisting of a
latent variable given by
\[ y^{\ast} = \bx' \bbeta + \ep, 
\]
where
\begin{equation} \ep \mid \bz \sim \mathcal{N}\left(0, \exp( 2\bz'\bgamma )
  \right). \label{stddevgamma}
\end{equation}
But one can also think of it as a model with a latent variable with mean
$\dspl{\frac{\bx' \bbeta}{\exp( \bz'\bgamma )}}$ and a disturbance
term distributed as a standard normal variable. We also observe that
there can be conditions in which the model is not identified, for
example if $\bz$ consists only of a constant.

One can estimate this model using maximum likelihood, much as one does
for the probit model. Assume that there are $n$ observations of $y_i$,
$\bx_i$ and $\bz_i$. For convenience, let $\by$ be the $n \times 1$
vector of all observations of $y_i$, let $X$ be the $n \times k_1$
matrix whose $i^\ith$ row is $\bx_i'$ and let $Z$ be the $n \times k_2$
matrix with $i^\ith$ row given by $\bz_i'$. 
Then the log likelihood function is given by:
\begin{equation} \ell \left(\bbeta,\bgamma \mid \by, X, Z\right) = \sum_{i=1}^n \left( y_i \ln \Phi
   \left( \frac{\bx_i' \bbeta}{\exp( \bz_i'\bgamma )} \right) + (1- y_i)
   \ln \left( 1 - \Phi
   \left( \frac{\bx_i' \bbeta}{\exp( \bz_i'\bgamma )} \right) \right)
  \right). \label{loglikelihood}
\end{equation}
Given a set of $n$ observed values of the random vectors $y,\bx$ and
$\bz$, in order to obtain the maximum likelihood estimate, one
maximizes this function over the space of possible choices of
$(\bbeta,\bgamma) \in \bbR^{k_1+k_2}$.

\section{Search Algorithm Performance}

\label{sec:performance}

We now examine how standard optimization algorithms perform when
maximizing the heteroskedastic probit log likelihood function.

\subsection{Simulated Data: Heteroskedastic Probit Model}

\label{subsec:simulateddata}

We first consider the behavior of search techniques when employed on
simulated data sets. We note that this and other data
sets used in the paper are not designed to be pathological. 
In this section, we describe the simulated data generating
process and then use four different search methods on this one data set. 

The basic data generating process is as follows. First, we
generate 1000 observations of $\bx_i$ and $\bz_i$, with components equal
to $0$ or $1$. Then we generate a particular choice for both $\bbeta$ and $\bgamma$ that
we call the {\it model parameters}, which we denote by $\bbeta_0$ and
$\bgamma_0$. Given these choices, for each observation $i$, we generate a
disturbance term $\ep_i$ sampled from the normal distribution with mean
$0$ and variance $\dspl{\exp(\bz_i' \bgamma_0)}$. Then we consider each sum
\begin{equation} 
\bx_i' \bbeta_0 + \ep_i. \label{eq:dgpeq1}
\end{equation}
If this sum is greater than or equal to $0$, we define $y_i$ to be $1$,
and we define $y_i$ to be $0$ if the sum is less than $0$. In this
fashion, we generate the parameters $\bbeta_0$ and
$\bgamma_0$, and the data set $\{ \bx_i,\bz_i, y_i \ ; \ \ 1 \leq i \leq
1000\}$. 

Now consider one potential complication. Suppose that for all $i$, the
term $\bz_i' \bgamma_0$ is large and negative, so that
the variance $\dspl{\exp(\bz_i' \bgamma_0)}$ is quite small. In this
case, it could be that the term $\bx_i' \bbeta_0$ in the sum
(\ref{eq:dgpeq1}) above is much larger in magnitude than that of the term $\ep_i$ for
all observations, and thus may completely determine the value of
$y_i$. We refer to this as a situation in which there is no {\it crossover}. Now define $\mathbb{1}(v)=1$ for $v \geq 0$ and $0$ for $v<0$. Then, more formally, we say that there is crossover for observation $i$ if
\[  y_i \neq {\mathbb{1}}\left( \bx_i' \bbeta_0 \right) ,
\]
that is, if the disturbance term leads to $y_i$ having a different value
from that which the linear term $\bx_i' \bbeta_0$ alone would predict. If all of
the observations exhibit no crossover, then it seems intuitively likely
that there will be no optimal choice
for $\bgamma$; the basic idea is the following.\footnote{This is of
  course exactly similar to the complete separation problem for the probit
  model. For one discussion of the complete separation problem, see \cite{Albert:1984}. Separately, we note
    that our purpose here is not to provide a complete
    proof, but only to explain the reasoning behind the choices in our
    data generating process.} Suppose, to the contrary, that there is an
optimizing argument and suppose for example that it occurs
  when $\bbeta=\bbeta_0$. For those observations
with either $\bx_i' \bbeta_0>0$ or $\bx_i' \bbeta_0<0$, because the components
of $\bz$ are nonnegative, one can observe from Equation (\ref{loglikelihood})
that the summands of the log likelihood function will generally increase
as the components of $\bgamma$ grow large and negative.\footnote{We
  assume here for simplicity that many of the components of the $\bz_i$
  terms are strictly positive.} For this reason, and
more importantly to focus on the class of models for which the
variance portion of the model plays an important role, we ensure that in
the data generating process there is a significant percentage
of crossover.

We now turn to the details of simulating the data. We take the dimension
of $\bx$ to be three and of $\bz$ to be two. The first component of $\bx$
is a constant equal to one, while the second and third components, as
well as the two components of $\bz$, are realizations of Bernoulli
variables with probability $\sol{1}{2}$, i.e. discrete uniform $\{ 0, 1
\}$ random variables. We generate 1000 such observations $\bx_i$ and
$\bz_i$. As mentioned above, we refer to the parameters used in the data generating process
as the {\it model parameters} and denote them by $\bbeta_0$ and
$\bgamma_0$. The components of these parameters are themselves sampled
from a (continuous) uniform distribution on $[-5,5],$ although there are
important provisos to this statement. For one, we select the second and
third component of $\bbeta_0$ from $[-5,5]$, but after doing so we in
fact choose the first component so that the mean of $\bx_i' \bbeta_0$ over all observations is $0$. The purpose of
this is to avoid the situation in which $\bx_i' \bbeta_0$ is, say,
positive for all or most of the data set; unless the variance is fairly
large, we might expect to see very little crossover. The second caveat
to the statement that the parameters are sampled from a uniform
distribution is that we discard any choices $(\bbeta_0,\bgamma_0)$ of
the parameters for which the crossover is less than 20\% or greater than
30\%; i.e., we simply repeat the random sampling until we find such a choice
of parameters.\footnote{The percentages chosen here are somewhat arbitrary, but
  reflect a desire for the average summand in the log likelihood
  function to be somewhat better (less negative) but not too much better than what
  turns out to be an essentially minimal value of $\log
  (\sol{1}{2})$. If there is too much crossover, this average summand (which we
  later refer to as the normalized value of the log likelihood function)
  will generally be very close to $\log (\sol{1}{2})$, whereas if there
  is very little crossover, it will generally be much better. In the
  former case, the value of the log likelihood at the model parameters
  is not much different from its value at what we call below the plateau
  solution, so it is hard to demonstrate the difference between the
  two. In the latter case, the model approaches a condition much like
  the case of complete separation in the probit model, so we want to
  avoid this situation. Recall as well that our purpose in this paper is
  merely to demonstrate that under reasonable conditions, there are
  concerns with optimization when using the heteroskedastic probit
  model, as opposed to showing that there are concerns under any
  conditions. Also, we note that we discuss why the value $\log
  (\sol{1}{2})$ arises more below.} Having selected
$\bbeta_0$ and $\bgamma_0$, then for every observation indexed by $i$, we sample a disturbance term $\ep_i$ from a (pseudo-)normal
distribution with mean $0$ and variance $\dspl{\exp(\bz_i' \bgamma_0)}.$
Then we define
\[  y_i = \mathbb{1}\left( \bx_i' \bbeta_0 + \ep_i \right) .
\]
This gives the complete set of simulated data $\{ \bx_i,\bz_i, y_i \  ; \ \ 1 \leq i \leq
1000\}$. Now we fix one data set simulated in this fashion, and consider the
performance of different algorithms when maximizing the log
likelihood on this data set.\footnote{The model parameters for this data set, to two decimal places, are given by $\bbeta_0=(.56,-.31,-.84)$ and $\bgamma_0=(3.43,-4.07).$} 


We first present results for the algorithm commonly referred to as the BFGS algorithm \citep{Broyden:1970, Fletcher:1970,Fletcher:1964,Shanno:1970}. In each of 1000 runs, we randomly select a starting value for the parameter
$(\bbeta_0, \bgamma_0)$ to be estimated. Each choice of initial values has
components in the interval $[-5,5]$, so that each choice of initial values is in the box $\dspl{[-5,5]^{5}}$. Denote the estimate resulting from the $i^\ith$ run by
\[ \left( \widehat{\bbeta}_{BFGS,i}, \widehat{\bgamma}_{BFGS,i} \right) ,
\]
and the set of all such estimates by
\[ E_{BFGS} = \left\{ \left( \widehat{\bbeta}_{BFGS,i},
    \widehat{\bgamma}_{BFGS,i} \right),  \ \  1
  \leq i \leq 1000 \right\} .
\]

We are interested in how close the estimates are to the model
parameters. We employ two measures to measure the proximity of the estimates in
$E_{BFGS}$ to the model parameter values. The first measure is the simple
Euclidean distance between each estimated parameter value and the model
parameter value. The second measure starts by looking at the difference
between the value of the log likelihood function $F$ at the model
parameter vector and the value of the log likelihood function $F$ at each
estimated parameter vector. We divide this difference by the number of observations in the data
set. The idea here is to normalize the difference in order to better
compare results using this data set with later results using a real data
set, which has a different sample size. Here the number of
observations is $1000$, which we denote by $n_{SIM}$ for expositional
purposes.


We present histograms of these two measures over all 1000
runs. We look at two histograms, for frequencies observed in
each of the following two sets, which correspond to the two measures
described above:
\[ D_{BFGS} = \left\{ \left\|
    (\widehat{\bbeta}_{BFGS,i},\widehat{\bgamma}_{BFGS,i}) -
    (\bbeta_{0},\bgamma_{0)} \right\|_2 ,\  1
  \leq i \leq 1000 \right\},
\]
where $\| \ \ \|_2$ denotes Euclidean distance, and
\[ V_{BFGS} = \left\{ \frac{1}{n_{SIM}} \left[
    F((\bbeta_{0},\bgamma_{0})) -
    F((\widehat{\bbeta}_{BFGS,i},\widehat{\bgamma}_{BFGS,i})) \right] , \ 1
  \leq i \leq 1000 \right\} .
\]
We note that $\dspl{\frac{1}{n_{SIM}} F((\bbeta_{0},\bgamma_{0}))= -0.45946.}$ There
is in fact an added complication for the histogram for $V_{BFGS}$, which
we discuss momentarily.

\begin{figure*}[!h]
\centering
\includegraphics[width=0.98\textwidth]{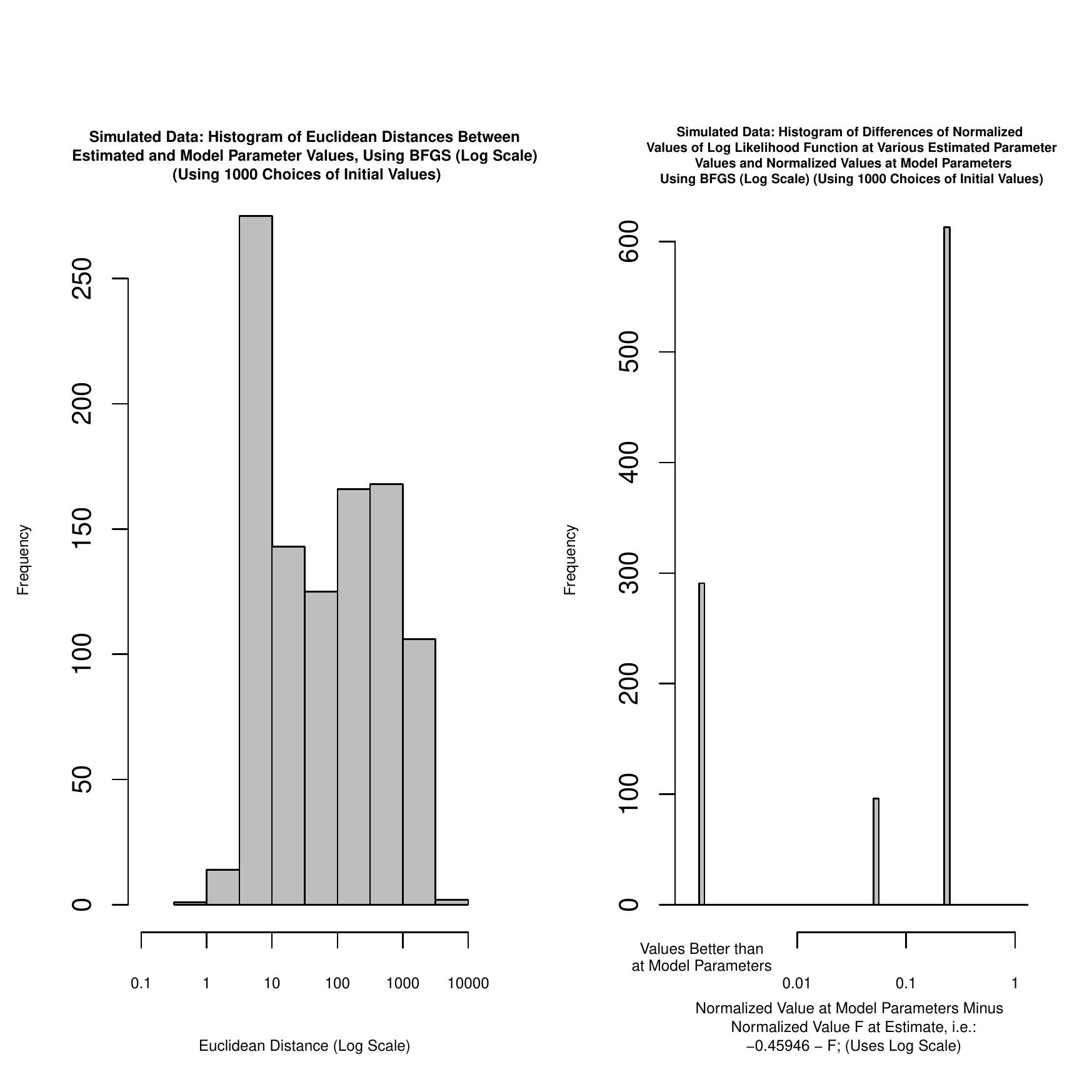}
\caption{Performance of the BFGS Search Algorithm on a Simulated Data
  Set for the Heteroskedastic Probit Model, for 1000 Random Choices of Initial Values}
\label{fig:simdata1}
\end{figure*}

The two histograms are displayed in Figure \ref{fig:simdata1}. The left
histogram plots frequencies for the set $D_{BFGS}$, and the right
histogram plots frequencies for the set $V_{BFGS}$. The left plot uses a log scale on the
horizontal axis, to more clearly present key phenomena. The right plot also uses a log scale on the horizontal axis, but only in the right section of the graph. There are actually almost 300 choices of starting values for which the log likelihood values at the estimates found using the BFGS algorithm are actually greater (i.e., less negative) than the value at the model parameters. This is not that surprising-- we are using a finite sample, and obviously not a population. For the log likelihood function, which
is the empirical analogue of the expectation of a generic summand over
the population, we do not have any reason to believe that the {\it log
likelihood} should be maximized at the model parameters; on the
other hand, under suitable conditions we might expect that the
{\it expectation} would be maximized at the model parameters.\footnote{See,
  for example, Lemma 14.1 of \citet{Ruud:2000}, for a statement of a result of
  this type concerning the expectation.}


Naturally, these negative differences would not show up on
the graph if we were to use a log scale, given that the log scale would
only allow representation of those observations for which the log
likelihood has a value at the estimated parameters that is in fact
strictly less than the value of the log likelihood at the model parameters. We represent the
observations for which the value of the log likelihood at the estimated
parameters is actually greater than the value of the log likelihood at
the model parameters in the graph in the following way. We plot a rectangle on the left side of the plot with height equal to the number of these cases and labeled as ``Values Better than
at Model Parameters.''\footnote{Its location along the horizontal axis is
  of course not meaningful.}  As a final note concerning the two plots
in Figure \ref{fig:simdata1}, observe that the scales for the
horizontal and vertical axes differ by plot.

When starting with random initial values, the parameter estimates do not
generally appear to be close to the model parameters, 
for about 70\% of the 1000 runs. In fact, the Euclidean distance between the
parameter estimates and the model parameters is at least $\dspl{3.16 \
\left(=10^{.5} \right)}$ for all but $15$ of
the $1000$ runs and at least $10$ for $790$ of the runs. Moreover, for
more than 600 runs, the value of the log likelihood function at the estimated
parameter values is at least $223.87$ less than the value at
the model parameters, which is $-459.46$.\footnote{The value $.22387$ equals
  $\dspl{10^{-0.65}}$, which is the left endpoint of the rightmost
  histogram cell in the plot on the right in
  Figure \ref{fig:simdata1}; recall that this value was obtained by normalizing, i.e. by
  dividing by $1000$.} Thus by two measures the BFGS algorithm appears
to only perform well for some fraction of starting values chosen
randomly from a box, at least for this data set.  
%

\begin{figure*}[!h]
\centering
\includegraphics[width=0.98\textwidth]{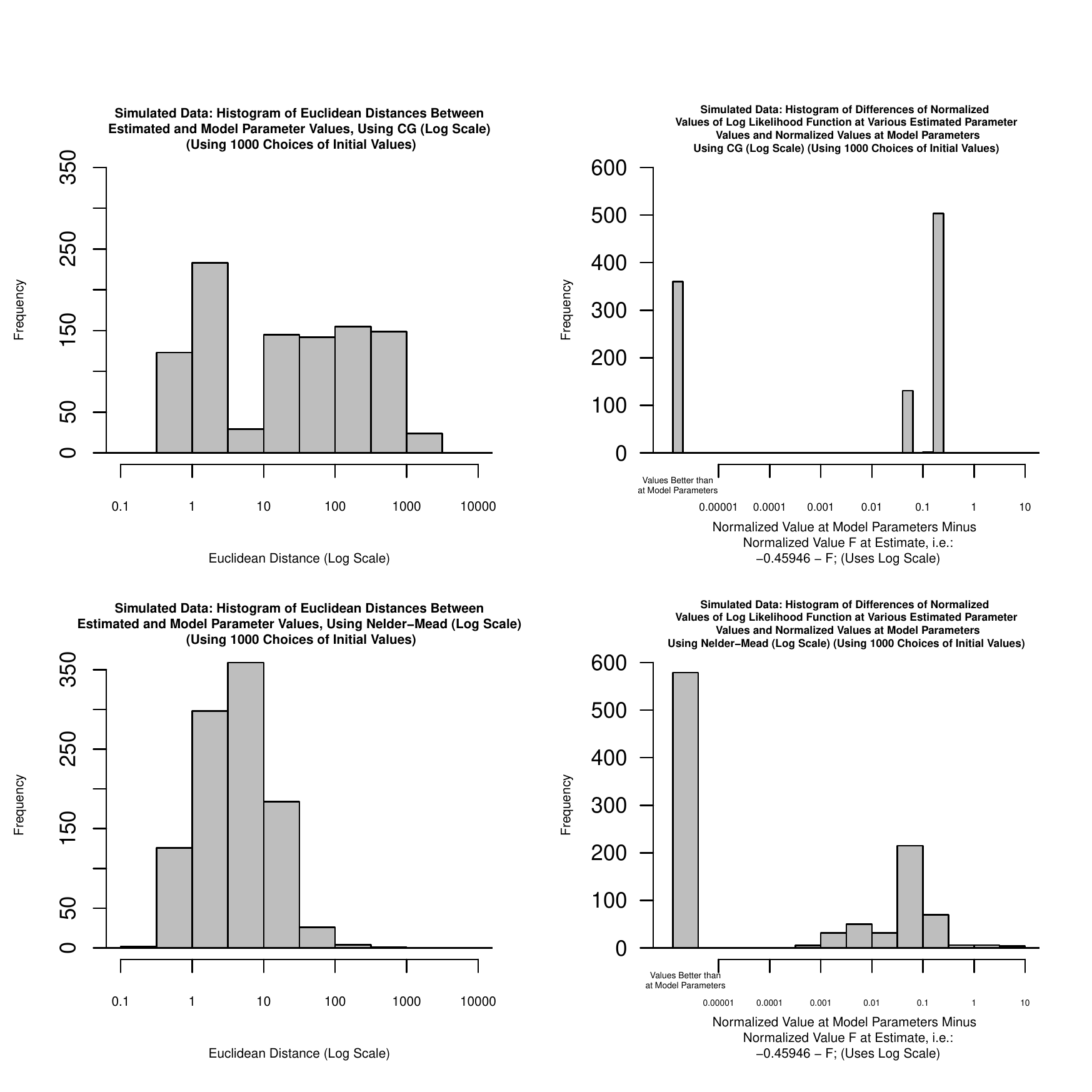}
\caption{Performance of the CG and Nelder-Mead Search Algorithms on a
    Simulated Data Set for the Heteroskedastic Probit Model, for 1000 Random Choices of Initial Values}
\label{fig:simdata2}
\end{figure*}

Figure \ref{fig:simdata2} presents similar plots for the same simulated
data set, but with two other optimization techniques. The simulations were
run so that the same 1000 choices of initial parameter vector were made
for each search technique. (To be clear, these are also the same initial choices
used for the BFGS search technique.) We combine results for two separate algorithms in one
figure. The top two plots present measures of the quality of estimates
found from implementing a conjugate gradients method (referred to as CG)
developed by \citet{Fletcher:1964}, while the bottom two plots display results found using a method developed by \citet{Nelder:1965}. For these figures, we adjust the plots so that the horizontal and vertical scales are the same within each column (but not within each row).

The first row considers the CG
method. Looking only at the right histogram, we see that at least half
of the initial values result in parameter estimates that give log
likelihood values more than $100 \ (= .1 \times 1000)$ less than the log
likelihood value at the model parameters. The second row reports on the
performance of the Nelder-Mead algorithm. It actually performs fairly
well, but the right histogram shows that there are roughly 300
starting values at which the log likelihood is more than $\dspl{31.6 \
\left(= 10^{-1.5} \times 1000\right)}$ less than the log likelihood
value at the model parameters.

We also implemented a simulated annealing algorithm, specifically a variant given
by \citet{Belisle:1992}, which we refer to in the paper as SANN. Due to
its less common usage in political science, we present those results in the
appendix. (See Figure \ref{fig:simdata3}.) Briefly, the SANN algorithm
performs very well, with all but $41$ of the $1000$ initial values
leading to estimates for which the log likelihood values are greater
than the log likelihood values at the model parameters. 


In addition, we looked at a few alternative situations.For one, we
  considered the case of continuous values of the data for the $\bz$
  variables, rather than the discrete values used in the simulated
  data set discussed earlier. (We still take the components of $\bz$ to
  be nonnegative.) Running search algorithms on a simulated
  data set with continuous $z$ variables led to problems like those found
  above, namely estimated parameters far from the model parameters, and
  values of the log likelihood function at the estimated parameters far
  from its values at the model parameters.\footnote{We note that
    our intuitive explanation given in Section
    \ref{subsec:intuitionsubsec1} indicates that these
  problems would arise for both discrete and continuous $\bz$. In fact,
that reasoning might lead us to believe the problems could be even worse for
continuous $\bz$, given that we would expect far fewer occurrences of
$\bz=\bzero$ in the continuous case.}


  We also looked at some simulations in which the model parameters
  $\bgamma$ were a bit smaller in absolute value. Recall that the model parameter
  $\bgamma_0$ in
  the data set above is $(3.43, -4.07)$. The reason for doing so is that
  one could be concerned that this value would lead to a large variance
  (exp($3.43$)) in the case that $\bz=(1,0)$ and a small variance
  (exp($-4.07$)) in the case that
  $\bz=(0,1)$; on the other hand, in the case that $\bz =(1,1)$, the variance
  (exp($3.43 -4.07$)) would not be as large or small. There were similar problems with the
  search algorithms for a data set with
  $\bgamma_0=(-0.6,0.84,-0.69,-0.15,-0.16,0.42),$ although the problems
  did occur with less frequency (as a fraction of the number of choices of initial values).\footnote{We note that, in very
    limited testing, we did
  not find any problems with local search for simulated data sets with
  $\dim(\bgamma_0)=2$.} We also note that below we find similar
  problems with a real data set used by Alvarez and Brehm, which has
  a parameter $\bgamma$ whose components are each smaller than 1 in
  absolute value.\footnote{These are not exactly a model parameters in
    this case, but is analogous-- see below.} 


\subsection{Simulated Data: Probit Model}
\label{subsec:simulateddataprobit}

We now contrast the results for the above search algorithms for the
heteroskedastic probit model with their performance for a probit model
with the same number of parameters.

Thus, we consider a probit model with $\dim(\bbeta)=5$, so that the dimension of the parameter vector is the same in both cases. We again generate a pseudorandom data set with 1000 observations $\bx_i$, with components equal to $0$ or $1$ (and with the first component always equal to $1$). Then we
generate a choice of $\bbeta_0$ by selecting the second through fifth
components of $\bbeta_0$ from $[-5,5]$, but we choose the first
component so that the mean of $\bx_i' \bbeta_0$ over all observations is $0$. In the probit
setting, we do not make any requirements about the fraction of
crossover. We again generate $1000$ choices for initial parameters in the box
$\dspl{[-5,5]^5}$ and start each search technique at all $1000$ of these
choices.

\begin{figure*}[!h]
\centering
\includegraphics[width=0.98\textwidth]{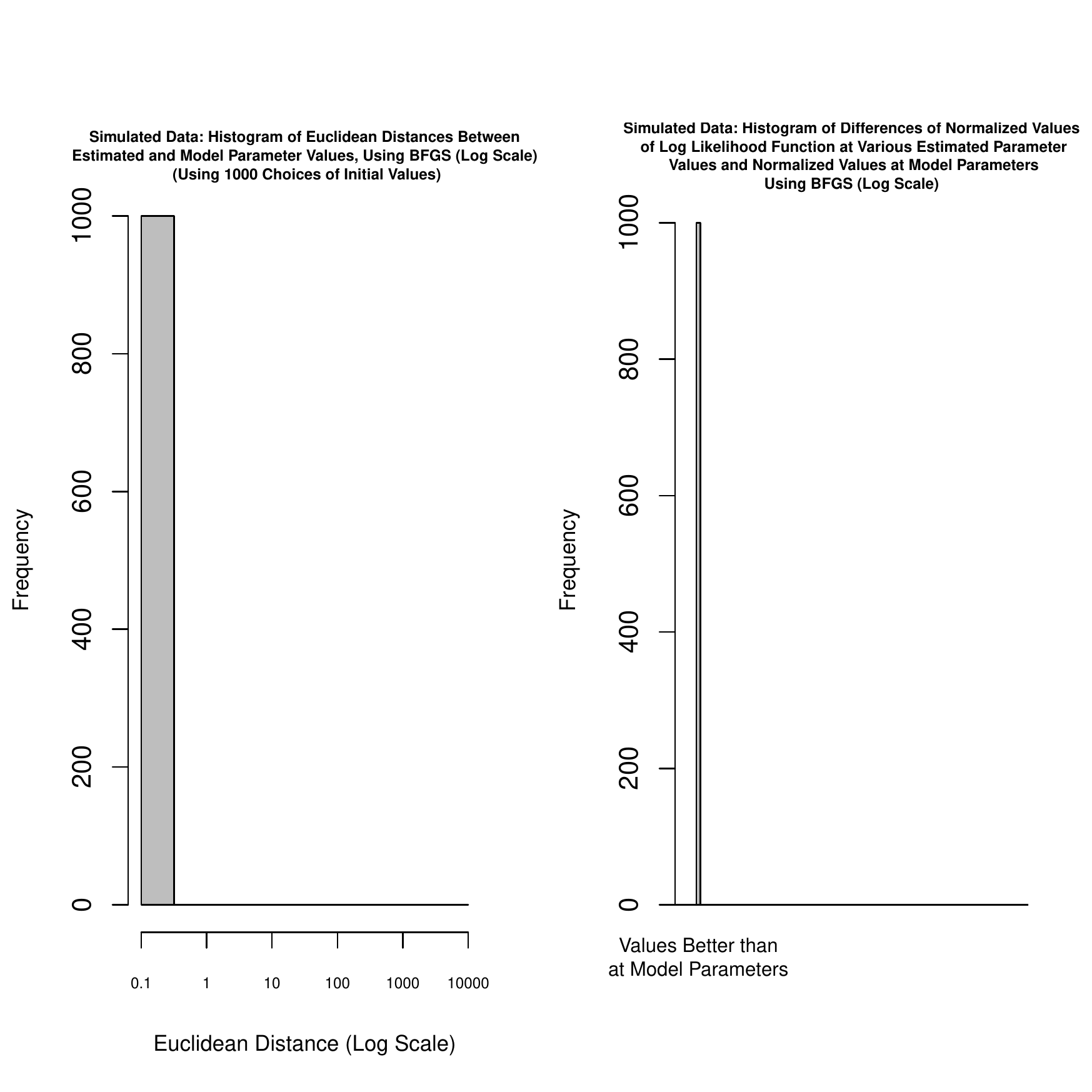}
\caption{Performance of the BFGS Search Algorithm on a Simulated Data
  Set for the Probit Model, for 1000 Random Choices of Initial Values}
\label{fig:simdataprobit1}
\end{figure*}

Results for the BFGS search algorithm are presented in Figure
\ref{fig:simdataprobit1}. All $1000$ initial values lead to estimates
at which the log likelihood is greater than at the model
parameters. Moreover, all of the estimates are in fact within $.233$ of
the model parameters.  These graphs can be compared directly with Figure $\ref{fig:simdata1}$.

\begin{figure*}[!h]
\centering
\includegraphics[width=0.98\textwidth]{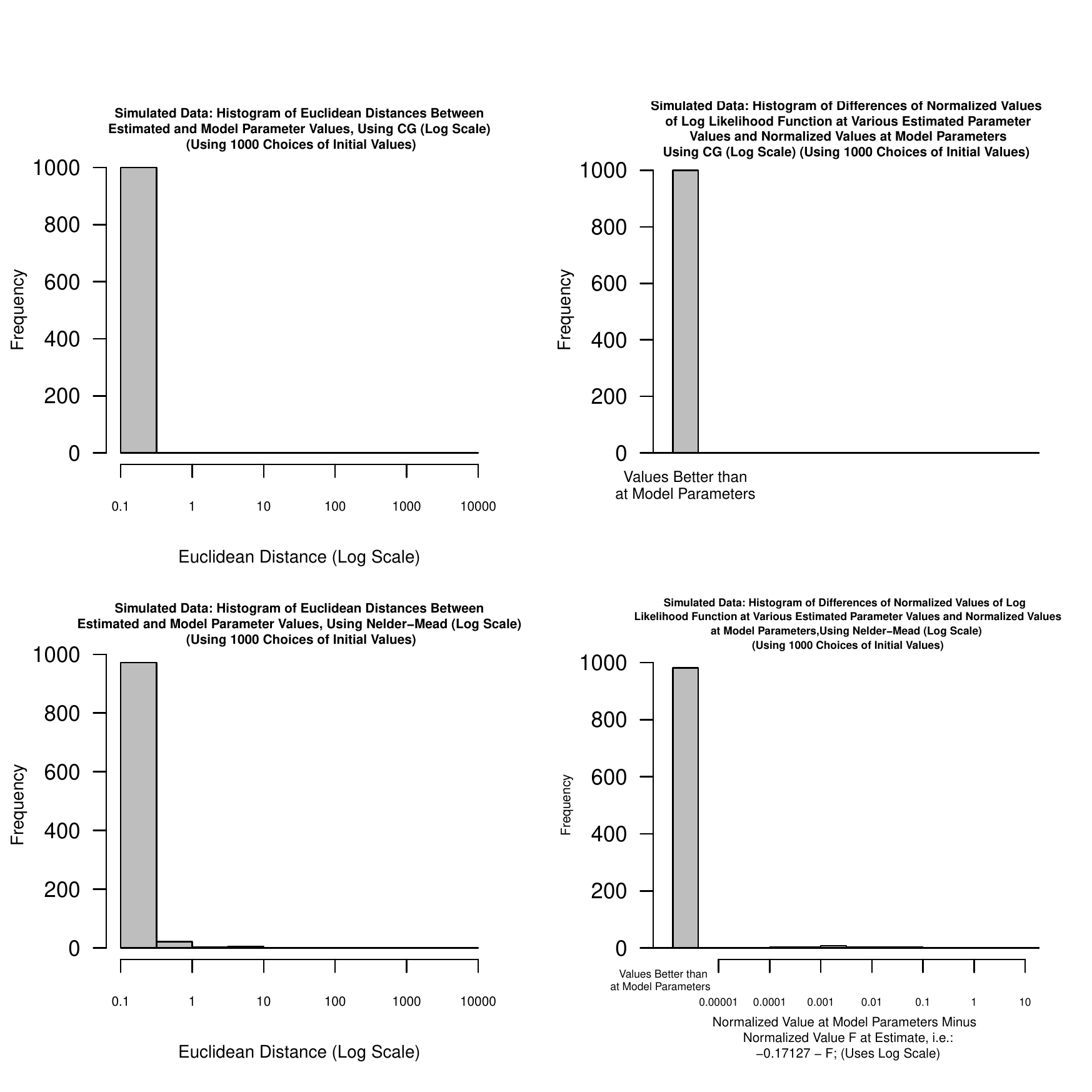}
\caption{Performance of the CG and Nelder-Mead Search Algorithms on a
    Simulated Data Set for the Probit Model, for 1000 Random Choices of Initial Values}
\label{fig:simdataprobit2}
\end{figure*}

For the CG and Nelder-Mead search algorithms, the results are 
similar. We present the results in Figure \ref{fig:simdataprobit2},
which can be compared directly with Figure \ref{fig:simdata2}. For the
CG algorithm, as with the BFGS method, all $1000$ initial values lead to
estimates at which the log likelihood is greater than at the model
parameters. The Nelder-Mead algorithm leads to some results that are not as
good; however, all but 19 initial values lead to estimates at which the log likelihood is
better than at the model parameters.\footnote{The results for the SANN
  algorithm are presented in Figure \ref{fig:simdataprobit3}, in the
  Appendix.} The performance of these search algorithms when applied to the probit model
is far better than the analogous performance for the
heteroskedastic probit model.

\subsection{Alvarez-Brehm Data: Heteroskedastic Probit Model}

\label{subsec:abdata}

We now consider the behavior of search techniques when employed on a
data set used by \citet{Alvarez:1995b} in their study of attitudes
toward abortion. We use the same specification as Alvarez and Brehm,
namely a heteroskedastic probit model using a constant and seven
variables in what they call the ``choice model'' (informally, the part depending on
$\bbeta$, so that the dimension of $\bbeta$ is eight), and six variables in the ``variance model''
(loosely, the part depending on $\bgamma$, so that the dimension of $\bgamma$ is six). Alvarez and Brehm actually use these explanatory variables in seven models, one for each of seven outcome
variables. Here we focus on one of these outcomes: whether or not
participants think that abortion should be legal if a woman wants it for
any reason.\footnote{See \citet{Alvarez:1995b} for more details. See Table 1 of their paper for results for the seven models.} In the data, we remove any observations for which one of the
explanatory or outcome variables is missing; in so doing, we match
Alvarez and Brehm in using 1295 observations.

\begin{figure*}[!h]
\centering
\includegraphics[width=0.98\textwidth]{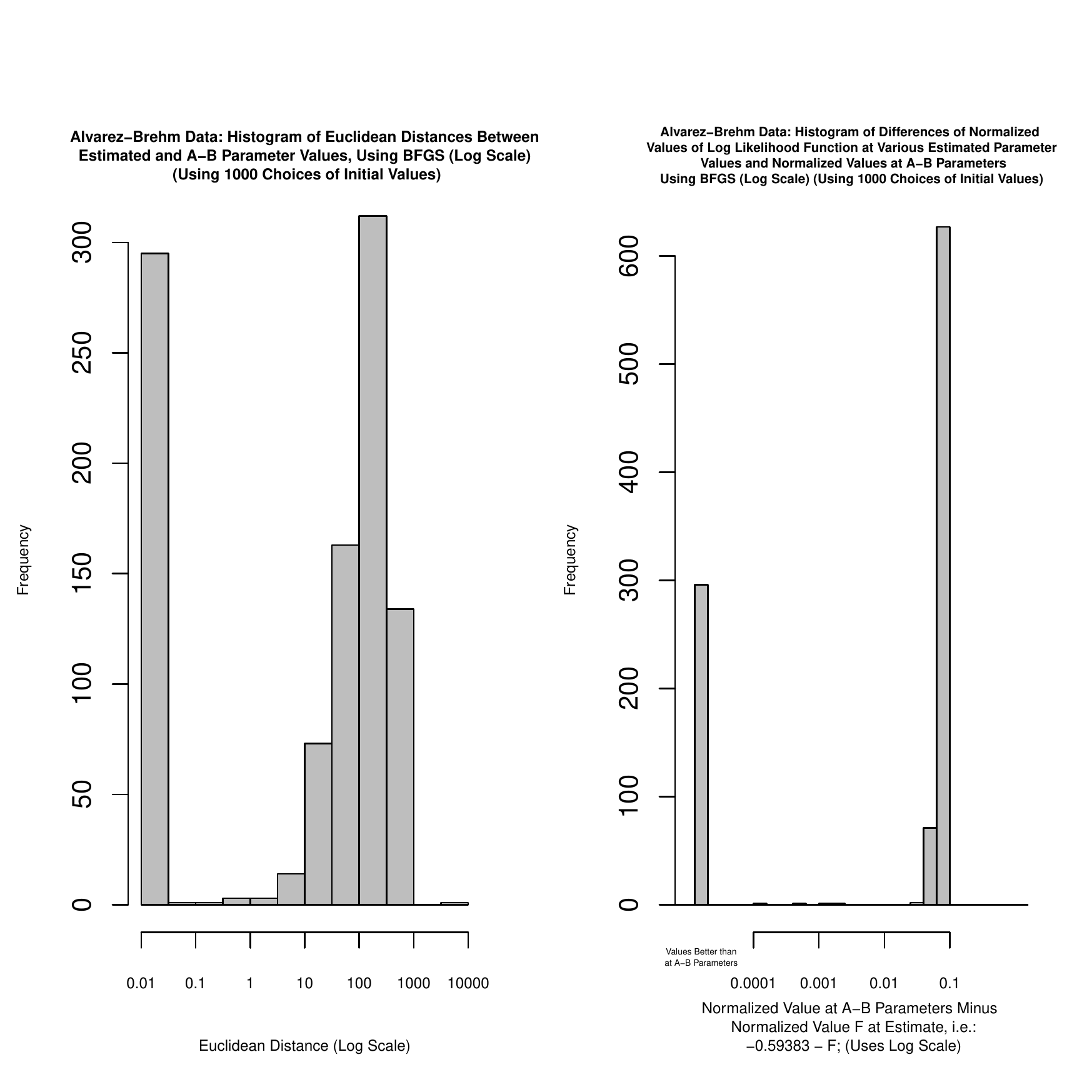}
\caption{Performance of the BFGS Search Algorithm on the
    Alvarez-Brehm Data (Using the Heteroskedastic Probit Model), for
    1000 Random Choices of Initial Values} 
\label{fig:abdata2}
\end{figure*}

As above, we are interested in how close the estimates are to optimal in some
sense. For operational purposes, we take the optimal value to be the
estimates found by Alvarez and Brehm, which we refer to as the A-B
parameter values. Denote this set of parameter values by
$\dspl{(\bbeta_{AB}, \bgamma_{AB})}$\footnote{These estimates can be
  found in the final column of Table 1 of their paper: they are
  $\dspl{ \bbeta_{AB}= (-.07, -.15, -.13, .05, -.22, -.79, .12, .51)}$
  and $\dspl{ \bgamma_{AB}= (-.22, -.48, .22, -.30, .68, .63)}$.}.
We now present histograms of the two measures used above to assess the
performance of the search methods. Here, for the normalized measure, we
will divide by $n_{AB}=1295$, the number of observations in the Alvarez-Brehm
data. We note that $\dspl{\frac{1}{n_{AB}}F((\bbeta_{AB},\bgamma_{AB}))= -0.59383.}$ 
We begin with the BFGS algorithm. The two histograms are displayed in Figure \ref{fig:abdata2}. Both plots
use a log scale on the horizontal axis, to more clearly present key
phenomena; as above, this log scale only applies to the right portion of
the right plot.\footnote{Observe that, as with the graphs for the simulated
  data, the scales for the horizontal and vertical axes differ by plot.}

Recall that, viewing the Alvarez-Brehm parameters as correct in some sense, we expect the log likelihood to be in general greater (i.e. negative and having less
magnitude) at these parameters than at our estimates. However, for
roughly 300 choices of starting values, the values of the log likelihood
at the estimates found using the BFGS algorithm are actually greater than
the value at the Alvarez-Brehm parameters. This is most likely due to
the fact that the parameters we use as the Alvarez-Brehm parameters are
from their paper, and reported only to two digits.\footnote{Note that
  this is unlike the situation for the simulated data discussed above, as here we would
  expect that Alvarez and Brehm found the estimates that maximize the
  log likelihood function, rather than finding the estimates that
  maximize the analogous expectation,
  which would typically not be maximized at the model parameters used to
  generate a simulated data set.} Thus we are probably
essentially getting the same parameter estimates as they are, as casual inspection of
a few of the better estimates indicates. As above, we represent this in the
graph by plotting a rectangle on the left side of the plot with height
equal to the number of these cases and labeled as ``Values Better than
at A-B Parameters.''\footnote{Again, its location along the horizontal axis is
  of course not meaningful.} 

When starting with random initial values,
the parameter estimates only appear to be close to the A-B parameters
for about 30\% of the 1000 runs. In fact, the Euclidean distance between the
parameter estimates and the A-B parameters is at least $10$ for $683$ of
the 1000 runs. Moreover, the {\it normalized} value of the log likelihood
function at the estimated parameter values is at least $.063$ less than
the {\it normalized} value of the log likelihood function at the A-B
parameters for more than 600 of
the runs; thus the (unnormalized) value of the log likelihood function is at least
$81.5$ less than the (unnormalized) value at the A-B parameters for these
runs.\footnote{The value $.063$ equals
  $\dspl{10^{-1.2}}$, which is the left endpoint of the rightmost
  histogram cell in the plot on the right in
  Figure \ref{fig:abdata2}; the value $81.5$ is just less than the product $1295 \times
  .063$.} Thus by these two measures the BFGS algorithm appears to
  perform well for only a fraction of the randomly chosen starting values. 

\begin{figure*}[!h]
\centering
\includegraphics[width=0.98\textwidth]{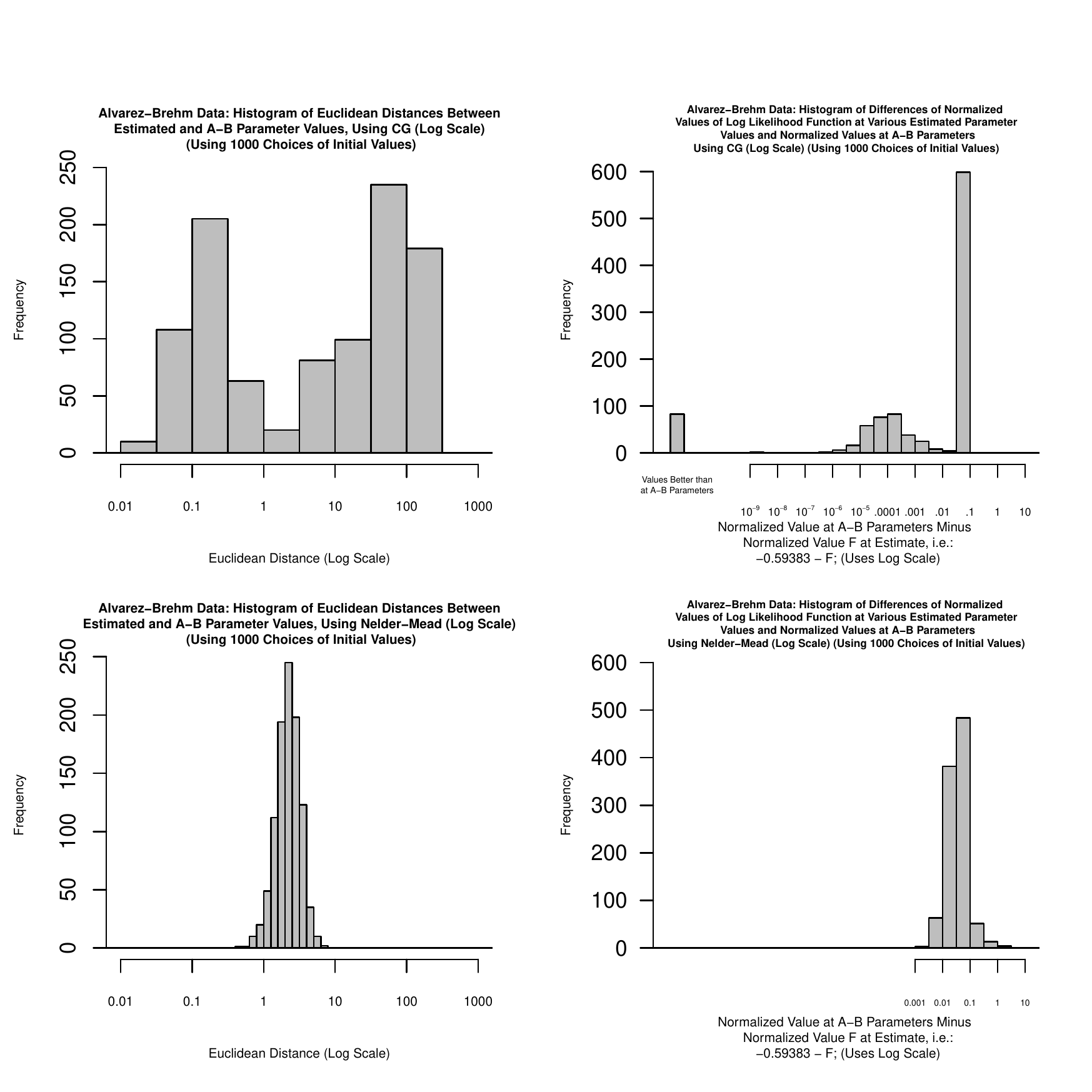}
\caption{Performance of the CG and Nelder-Mead Search Algorithms on the
    Alvarez-Brehm Data (Using the Heteroskedastic Probit Model), for
    1000 Random Choices of Initial Values} 
\label{fig:abdata3}
\end{figure*}

Figure \ref{fig:abdata3} presents similar plots, again for the Alvarez-Brehm
data set and starting at each of the same 1000 choices of initial values,
but now we look at the CG and Nelder-Mead algorithms. As above, for these
figures, we adjust the plots so that the horizontal and vertical scales
are the same within each column (but not within rows). 

From the right plot in the top row, one can see that at least roughly
600 of the results from the CG method appear to be fairly far from
optimal, as judged using the value of the log likelihood function at
these estimates. There are almost about 100 choices of initial value
however that yield estimates better than the A-B parameter values by
this measure, and a few that are only slightly worse (about
$\dspl{10^{-9}}$ or $\dspl{10^{-6}}$ larger). From the left plot in the
bottom row, all but about 100 of the estimated parameters from the
Nelder-Mead method have Euclidean
distance at least $1$ from the A-B parameter values; this may not
necessarily be indicative of poor estimates, but from the right
plot we can see that at least about $100$ estimates are not close, in that the
normalized values are at least $0.1$ larger than the normalized value at the A-B
parameters. (So the log likelihood function is at least $12.95$ larger.)
In the Appendix, we look at the performance of the SANN search algorithm
on the A-B data. (See Figure \ref{fig:abdata1}.) The results are neither
clearly better nor worse than those for the other search techniques.

\section{Observations on the Shape of the Log Likelihood Function for the Heteroskedastic Probit}
\label{sec:intuition}

\begin{figure*}[!h]
\centering
\caption{} 
\includegraphics[width=0.98\textwidth]{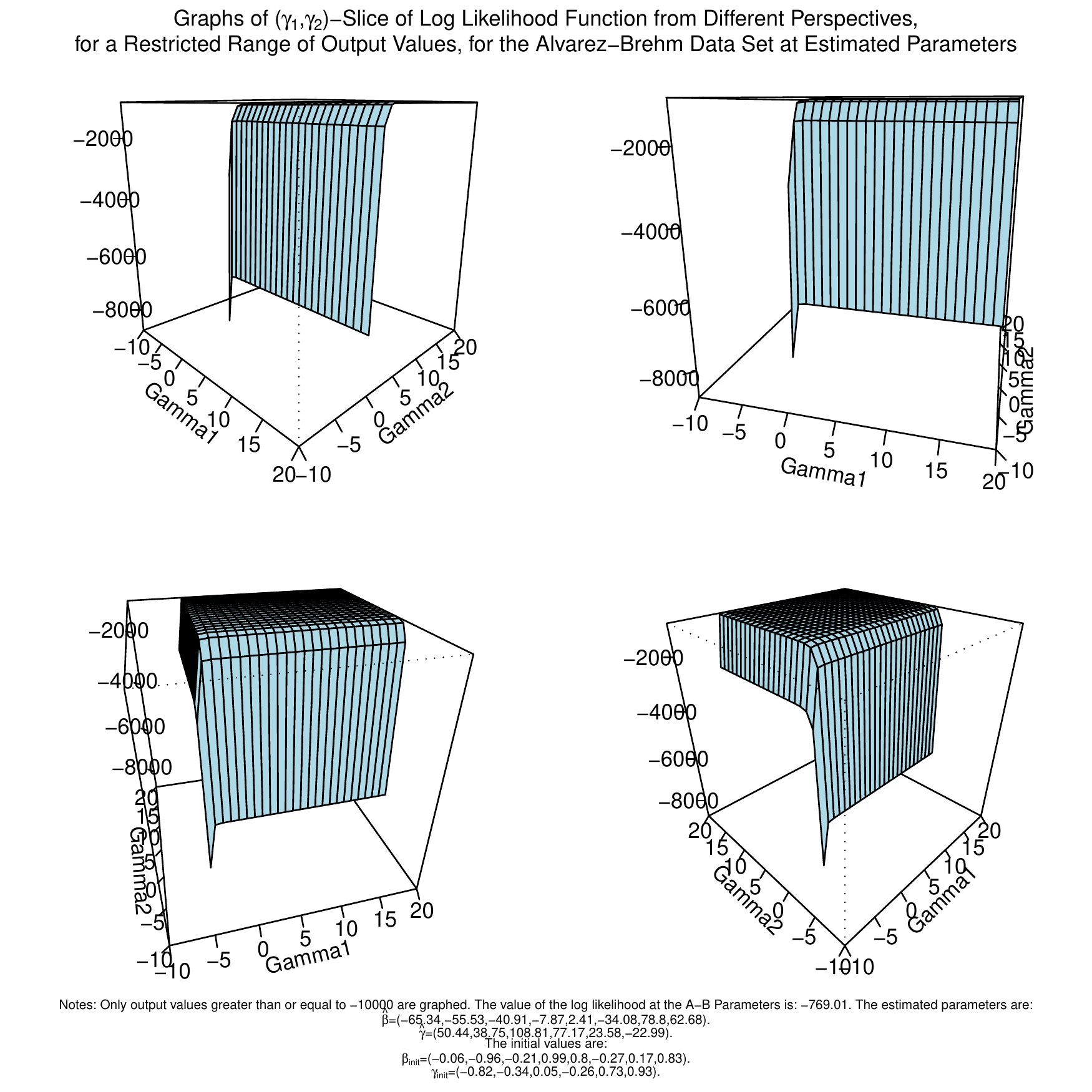}
\label{fig:abprofileBFGS}
\end{figure*}

Next, we explore why optimization algorithms might perform poorly when applied to the heteroskedastic probit likelihood.  We start our exploration with a profile plot of the log likelihood function for
the heteroskedastic probit using the the Alvarez-Brehm data. We first
fix a parameter estimate arising from the BFGS
algorithm for a particular initial value, which we have chosen to
have a resulting parameter estimate at which the value of the log
likelihood is not close to its value at the Alvarez-Brehm estimates. We display a graph that demonstrates how the log likelihood varies as the first two components $\gamma_1$ and $\gamma_2$ of
$\bgamma$ vary.\footnote{Throughout Section \ref{sec:intuition}, for ease
  of notation, we often use $\bbeta$ and $\bgamma$ where 
  $\dspl{\widehat{\beta}}$ and $\dspl{\widehat{\bgamma}}$ might perhaps
  be more appropriate.}  The value of
$\bbeta$ and the other components of $\bgamma$ are fixed at the estimated values. Figure \ref{fig:abprofileBFGS}
presents this plot, from four different vantage points. In order to make the graph
readable, we restrict the range of output values plotted to be those
values at least $-10,000$. At the values of $(\gamma_1,\gamma_2)$ for
which no output values are plotted, the log likelihood function is
actually much less than $-10,000$.

We see a distinct shape in this graph: a plateau, which ranges roughly
over positive values of $(\gamma_1,\gamma_2)$. This gives a sense of the plateau but 
investigation of the graph restricted to a smaller range of output values shows that it has a
more complex structure. For example, there is a ridge for $\gamma_2$
fixed at a certain value a little less than zero, which declines in
$\gamma_1$. If this plateau were in some sense flat enough, 
an optimization algorithm that is essentially local in nature may not be able
find the actual maximum of the function if it starts at an initial value
somewhere on the plateau. Additionally, it may be
led to this plateau from other regions and then subsequently remain
there. 

\subsection{The Heteroskedastic Probit Log Likelihood Function}

\label{subsec:intuitionsubsec1}

Next we discuss why we might expect a plateau region of the
sort seen in Figure \ref{fig:abprofileBFGS}. Consider the form of the log
likelihood function for the heteroskedastic probit, given in
Equation (\ref{loglikelihood}).

In this section, we focus on the case in which $\bz$ is discrete and,
for all $i$ with $1 \leq i \leq n,$ all of the components of $\bz_i$ are
nonnegative. For convenience, we denote this by the shorthand 
\[ Z \geq 0,
\]
where we are thinking of $Z$ as above, i.e. as the $n \times k_2$
matrix with $i^\ith$ row given by $\bz_i'$. In the Alvarez-Brehm
data used above, this holds, and it is also the case in the simulated
data in Section \ref{sec:performance}. We will return to consideration
of more general $\bz$ later. We will also assume in this section that
there are some values $i$ for which $\bz _i = \bzero$.
%

Then, except in the case that $\bz_i={\mathbf 0}$, for values of $\bgamma$
with components that are positive and sufficiently large, we can ensure
that $\bz_i' \bgamma$ is large and positive. Of course for such $i$, we also have that $\exp(
\bz_i'\bgamma )$ is large, whence the term $\dspl{\frac{\bx_i' \bbeta}{\exp(
  \bz_i'\bgamma )}}$ will be close to $0$. Thus,
under these conditions we have 
\begin{eqnarray} \ell \left(\bbeta,\bgamma \mid \by, X, Z \right) & \approx &
  \sum_{1 \leq i \leq n: \ \bz_i = 0} \left( y_i \ln \Phi
  \left( \bx_i' \bbeta \right) + (1- y_i)
  \ln \left( 1 - \Phi
  \left( \bx_i' \bbeta \right) \right)
\right) \nonumber  \\
&  & \hskip 40pt + \sum_{1 \leq i \leq n: \ \bz_i \neq 0} \left( y_i \ln
  \Phi(0)  + (1- y_i) \ln \left( 1 - \Phi(0) \right) \right) \nonumber \\ 
 & \approx &
  \sum_{1 \leq i \leq n: \ \bz_i = 0} \left( y_i \ln \Phi
  \left( \bx_i' \bbeta \right) + (1- y_i)
  \ln \left( 1 - \Phi
  \left( \bx_i' \bbeta \right) \right)
\right) \nonumber \\
&  & \hskip 40pt - (\ln 2) \times \bigg| \{\bz_i: \bz_i \neq \bzero, \quad 1 \leq i
  \leq n\} \bigg|.  \label{plateauform} \\ \nonumber 
\end{eqnarray}
From the form of the expression in \umpp{plateauform}, we can see why
there is a plateau in the graphs in Figure
\ref{fig:abprofileBFGS}. Consider a choice $\bgamma_1 = (\gamma_{11},
\gamma_{12}, \ldots, \gamma_{1k_2})$ with large positive
components. Throughout the remainder of the paper, we at times abuse language and refer to such vectors
$\bgamma$ informally as simply ``large and positive.'' Then, for
$\bgamma = (\gamma_{11}, \gamma_{12}, \ldots, \gamma_{1k_2})$ near
$\bgamma_1$, the denominator $e^{\bz_i' \bgamma}$ will still be large
and positive, so that $\dspl{\sol{\bx_i' \bbeta}{\exp( \bz_i'\bgamma)}}$
will be small and relatively unaffected by small changes in
$\bbeta$ and $\bgamma$. 


To see this in more detail, assume that there is a choice of
parameters $(\bbeta_1,\bgamma_1)$ and some large value of $L >0$ so that
$\bgamma_1$ satisfies 
\begin{equation} \min_{1 \leq j \leq k_2} \gamma_{1j} \geq L. \label{eq:gamma1large}
\end{equation}
Recall that we have assumed that the $\bz$'s are discrete and have
nonnegative components. Now assume for convenience that in fact the
smallest positive value of the components of the $\bz$'s is $1$.

Consider a new choice of parameters $(\bbeta_2,\bgamma_2)$ with
$\bgamma_2= (\gamma_{21}, \gamma_{22}, \ldots, \gamma_{2k_2})$
satisfying, say, 
\begin{equation} | \gamma_{2j} - \gamma_{1j} | \leq 1 \qfor 1 \leq j
  \leq k_2 . \label{eq:gammasclose}
\end{equation}
Then
\umpp{eq:gamma1large} and \umpp{eq:gammasclose} together imply that 
\[ \min_{1\leq j \leq k_2} \gamma_{2j} \geq L - 1.
\]
Also assume that $H>0$ satisfies
\[ |\bx_i' \bbeta_1 | \leq H \qand |\bx_i' \bbeta_2 | \leq H  \qfor 1
\leq i \leq n.
\]
Now we consider terms of the log likelihood function for which $\bz_i$
is not zero. Fix any $i_1$ with $1 \leq i_1 \leq n$ satisfying $\bz_{i_1} \neq \bzero$. 
As we have assumed that $\bz$ is discrete with a minimum nonzero value
of $1$, it follows that $\bz_{i_1}' \bgamma_2 \geq L - 1$, and thus
$\dspl{\exp({\bz_{i_1}' \bgamma_2}) \geq \exp(L - 1)}$ holds. Therefore one
has
\[ \left| \frac{\bx_{i_1}' \bbeta_j}{\exp({\bz_{i_1}' \bgamma_j})} \right| \leq
H e^{1-L} \quad \mbox{for} \ j=1 \ \mbox{and} \  j=2.
\]
So certainly for $L$ large relative to $H$, 
\[ \frac{\bx_{i_1}' \bbeta_1}{\exp({\bz_{i_1}' \bgamma_1})}  \qand
\frac{\bx_{i_1}' \bbeta_2}{\exp({\bz_{i_1}' \bgamma_2})}
\]
will both be small in magnitude, and thus both $\Phi$ and $(1-\Phi)$
evaluated at these expressions will be very close to $\sol{1}{2}$ (as
$\Phi(0)= \sol{1}{2}$). So the values of the $i_1^{\scriptsize{th}}$
summand of the log likelihood at the two parameter choices $(\bbeta_1,
\gamma_1)$ and $(\bbeta_2, \gamma_2)$ will clearly be very close to each
other; that is, the 
$i_1^\ith$ term of the log likelihood function will not change
substantially for small perturbations of $\bgamma$ with large, positive
components.  
Of course, for $i$ such that $\bz_i = \bzero$, the $i^\ith$
term of the log likelihood could change in a substantial way if $\gamma$
is on the plateau. But in
certain situations, these terms may be a small portion of the total
number of terms, and thus may not have a strong effect on the value of
the sum as a whole.

Now consider an optimization algorithm used to maximize the log
likelihood. Suppose the algorithm reaches a point $(\bbeta, \bgamma)$ on the plateau, i.e. with $\bgamma$
having large and positive components. As seen above, any small changes in
$\bgamma$, regardless of the change in $\bbeta$, will not substantially
affect the terms indexed by $i$ for which $\bz_i \neq 0$. We
propose that the effect of changes in the remaining terms might not be
enough to guide the algorithm to the best value; any
improvements in the terms indexed by those $i$ for which $\bz_i =
\bzero$ that might be caused by veering off of the plateau could perhaps
be outweighed by increased penalties in the far greater number of terms
indexed by $i$ such that $\bz_i \neq 0$.

Before turning to another topic, we make another observation. As discussed above, if $\bgamma$ has
large and positive components, then for $i$ with $\bz_i \neq \bzero$,
the expression $\dspl{\frac{\bx_{i}' \bbeta}{\exp({\bz_{i}' \bgamma})}}$
will be close to $0$. Thus $\dspl{\Phi \left(   \frac{\bx_{i}'
      \bbeta}{\exp({\bz_{i}' \bgamma})} \right) }$ and $\dspl{ \left( 1 - \Phi \left(   \frac{\bx_{i}'
      \bbeta}{\exp({\bz_{i}' \bgamma})} \right) \right) }$ will be close to
$\sol{1}{2}$. Hence the terms in the log likelihood indexed by $i$ for
which $\bz_i$ is nonzero will be close to $- \ln \ 2$. Thus, we expect to
see values of the log likelihood function close to $- n \ln \ 2$ at the
estimates on the plateau, especially if there are very few terms with
$\bz_i = \bzero$. Thus, in some sense, we can also think of $-n \ln \ 2$
as a sort of benchmark for a value that the optimization algorithm should be able to
easily achieve. Observe that $- \ln \ 2 \approx -0.693$. One can compare
the values obtained in the simulation to this value. For example,
consider the right plot in Figure \ref{fig:abdata2}. The normalized value of
the log likelihood function at the A-B parameters is $-.59383$, as can
be read off from the horizontal axis of the plot. There are more than
$600$ initial values from which the BFGS search method results in
estimates at which the normalized value of the log likelihood is almost
$.1$ less than the value at the A-B parameters, which gives a
normalized value on the order of $-\ln 2$. As we expect for
the reasons outlined here, it was in fact common
in our various computations for many of these normalized values to be
fairly close to $-\ln 2$ at plateau solutions.

Considering this benchmark gives another way to think intuitively about
the benefit to an optimizer of choosing a large positive value for $\bgamma$. The
function $\dspl{\ln \left( \Phi(x) \right)}$ decreases rapidly for $x$
negative and decreasing. (For example, $\dspl{\ln
  \left(\Phi(-10)\right)}\approx -53.$) So, in some sense, the penalty
for missing by a lot for one observation can be very large, whereas the
gain from improving the other observations may not be that large. So it
may be beneficial to simply make every term fit at least somewhat
decently, and a large positive value for $\bgamma$ allows us to at least
limit these penalties to around $-0.693$ each. This is obviously not a
rigorous line of reasoning, but it gives a heuristic for why search
techniques might often end with a plateau solution.

\subsubsection{Graphical Interpretation: The Basic Probit Model}

\begin{figure*}[!h]
\centering
\includegraphics[width=0.98\textwidth]{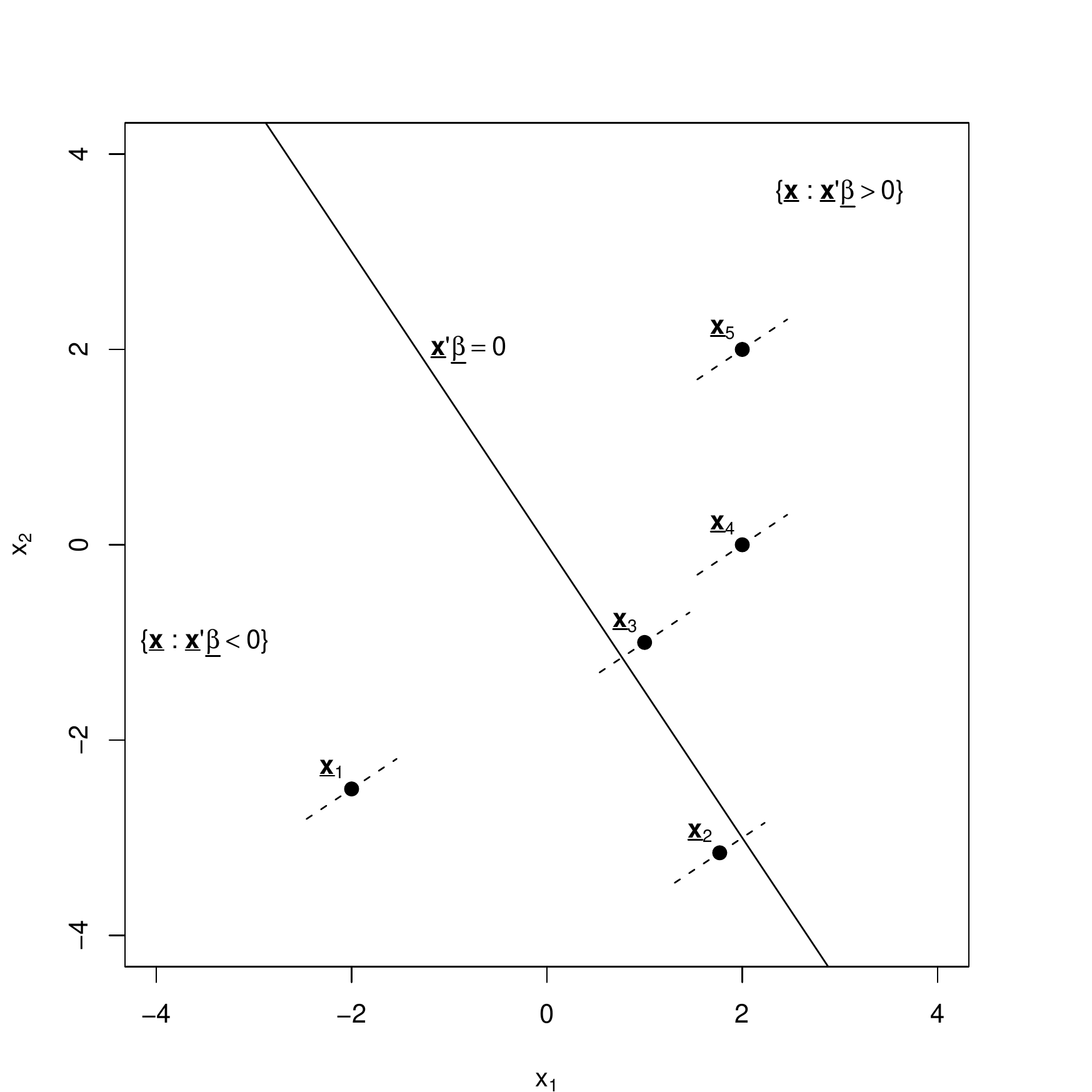}
\caption{A Graphical Representation of the Probit Model}
\label{fig:probitinterpret}
\end{figure*}

Now we give a graphical interpretation that sheds light on the
underlying mechanics of how optimization algorithms perform, when used to
maximize the log likelihood function associated with the heteroskedastic
probit model. To develop the interpretation, we first consider a probit
model with two-dimensional parameter $\bbeta$, so that $\mbox{Pr}(Y=1
\mid \bx) = \bx' 
\bbeta$. For a fixed choice $\bbeta \in \bbR^2$, Figure \ref{fig:probitinterpret} depicts several observations $\bx_1, \bx_2, \ldots, \bx_5 \in \bbR^2$ and the hyperplane $\bx' \bbeta = 0$, which of course in $\bbR^2$ is
simply a line. Around each observation $\bx_i$, we draw a dashed line segment
perpendicular to the line $\bx' \bbeta =0$, extending in either
direction from the point $\bx_i$; the endpoints are those points $\bv_{i1}$
and $\bv_{i2}$ on this perpendicular line that satisfy 
\[ \bv_{i1}' \bbeta = \bx_i' \bbeta -1 \qand \bv_{i2}' \bbeta = \bx_i'
\bbeta + 1. 
\]
This dashed line represents adding an error term $\ep_i$ to the linear
combination $\bx_i' \bbeta$. The values of $\bv' \bbeta$ for vectors
$\bv$ correspond to values of the latent variable $y_i^{\ast} = \bx_i'
\bbeta + \ep_i$ for $|\ep_i| \leq 1$.

The region $\{ \bx : \bx' \bbeta > 0 \}$ is, in this example, the upper
right half plane in Figure \ump{fig:probitinterpret}, because the
value of $\bbeta$ we use is $(1.5,1)$. This region roughly corresponds
to the region with those $\bx_i$ for which the associated $y_i$ equals
$1$, while the lower left half plane is the region $\{ \bx : \bx' \bbeta
< 0 \}$, which corresponds roughly to the region with $y_i = 0$; note
that we say ``roughly'' because one could have what we have termed
crossover.\footnote{We ignore points $\bx_i$ satisfying $\bx_i' \bbeta
  =0$ here as its consideration would only complicate the discussion,
  which we give only for the sake of intuitive understanding, and
  moreover would
  not add any material value.} Crossover occurs if either (i) $\bx_i '
\bbeta > 0$ holds yet $y_i =0$, or (ii) $\bx_i ' \bbeta < 0$ holds yet $y_i
=1$. For an example, consider $\bx_2$ in Figure
\ump{fig:probitinterpret}. One in fact has $\bx_2' \bbeta =
-0.5$. If we assume that $\ep_2 = 0.75$, then the latent
variable $y_2^{\ast} = \bx_2' \bbeta + \ep_2 = 0.25$, so $y_2=1$, and
thus for this value of $\ep_2$, observation $2$ exhibits crossover.

Now consider, in terms of Figure \ump{fig:probitinterpret}, what choices
might be made by an optimization algorithm used to maximize the value of
the log likelihood for the probit model. Despite the fact that generally
an algorithm is used to maximize the log likelihood, for exposition
purposes, we will generally speak of an ``optimizer'' as if an individual were
attempting to maximize the log likelihood. So consider the task facing
this optimizer. One key goal, given data $\{ (y_i, \bx_i) \}_{1 \leq i
  \leq n}$, is to choose $\bbeta$ so that most $\bx_i$'s are on
the correct side of $\bx_i' \bbeta = 0$, so to speak; i.e., one would
choose $\bx_i$ to be in the half-plane $\bx_i' \bbeta >0$ if $y_i =1$
and in the half-plane $\bx_i' \bbeta <0$ if $y_i=0$. In other words, an
optimizer would aim to minimize crossover. Of course, an optimizer need
not do so exactly.\footnote{And indeed it is probably extremely unlikely
  with real-world data to be able to do so, and if so, there would be
  the problem of complete separation.} We can think of each summand of
the log likelihood function as assigning a penalty if there is
crossover; of course, there is also a summand that is added even if
there is no crossover for the observation, but all of these summands are
negative and we think of most of them as having small magnitude, so that
intuitively, an optimizer might be mainly concerned with the summands
for which there is in fact crossover. These penalties for crossover are
determined as follows. First, consider what we have just noted: this penalty is always negative. For the points $\bx_i$ that are far from the line $\bx_i' \bbeta=0$, the penalty for
crossover in the log likelihood function is much larger (in magnitude) than the penalty
for points $\bx_i$ near the line. For example, the $i^\ith$ term
of the log likelihood function for $i$ with $y_i=1$ is $\ln \Phi (\bx_i'
\bbeta)$, which is of course much more negative, the more negative is the
value of $\bx_i' \bbeta$. So, for example, if $y_1=y_2=1$ were to hold,
then given their positions as shown in Figure \ump{fig:probitinterpret},
it is more important to put $\bx_1$ in the correct half plane than
$\bx_2$, as the penalty $\ln \Phi (\bx_1' \bbeta)$ from not doing so is
much larger (in magnitude) than the corresponding penalty $\ln \Phi (\bx_2'
\bbeta)$.


\begin{figure*}[!h]
\centering
\includegraphics[scale=0.5]{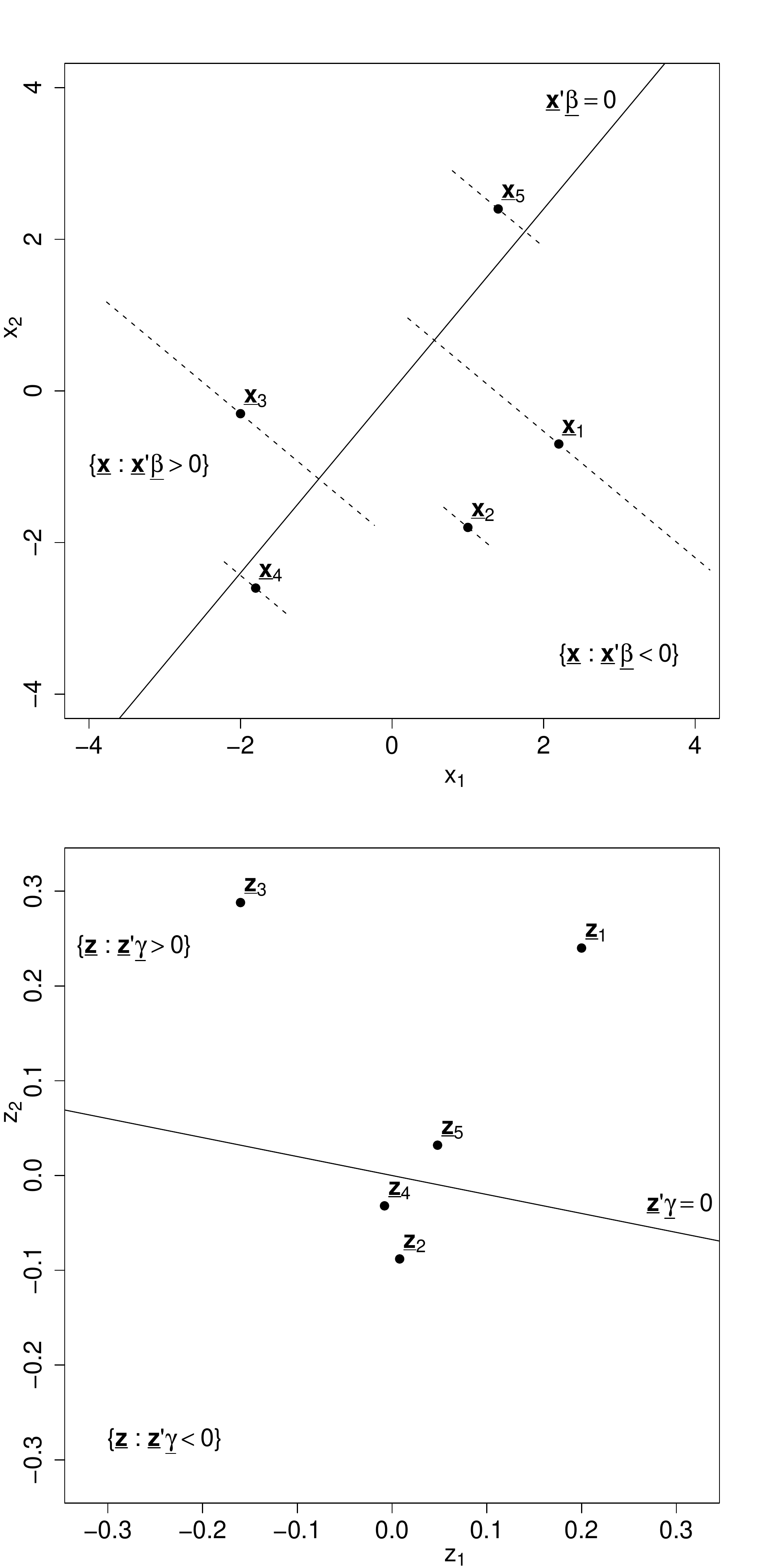}
\caption{A Graphical Representation of the Heteroskedastic Probit Model}
\label{fig:hetprobitinterpret}
\end{figure*}

\subsubsection{Graphical Interpretation: The Heteroskedastic Probit Model}

Now we consider the heteroskedastic probit model, and, for simplicity,
we examine the case in which $\dim(\bx)=2$ and $\dim(\bz)=2$, i.e. for
$\bbeta, \bgamma \in \bbR^2$. The picture here includes a few
modifications to Figure \ref{fig:probitinterpret}. Again, the objective
of an optimizer involves choosing $\bbeta$ to minimize crossover but now
with an additional tool, namely the optimizer can choose $\bgamma$ so as to
affect the penalties associated with crossover. Figure
\ref{fig:hetprobitinterpret} presents the graphs. The top graph
is similar to the plot in Figure \ref{fig:probitinterpret}. Again, the line
$\bx' \bbeta = 0$ separates the observations $\bx_1, \ldots, \bx_5$ into
two half-planes. Also, we again draw dashed line segments around each
$\bx_i$, in the direction perpendicular to the line $\bx' \bbeta =
0$. But now these line segments extend for lengths that vary for each
observation; this length, for the point $\bx_i$, varies with
$\exp(\bz_i' \bgamma)$; much as in the probit graph, the endpoints
of the $i^\ith$ line segment are the points $\bv_{i1}$ and $\bv_{i2}$
satisfying 
\begin{equation} \bv_{i1}' \bbeta = \bx_i' \bbeta - \exp(\bz_i' \bgamma) \qand
\bv_{i2}' \bbeta =  \bx_i' \bbeta + \exp(\bz_i' \bgamma). \label{vdefs}
\end{equation}
Recall from \umpp{stddevgamma} that $\exp(\bz_i' \bgamma)$ is the
standard deviation\footnote{The decision to use the
  standard deviation is somewhat arbitrary; we are merely showing
  one pictorial representation of the main concepts.} of the error term
$\ep_i$ that forms part of the definition of the latent variable
$y_i^{\ast}.$ For those observations $i$ with $\exp(\bz_i' \bgamma)$
large, speaking very loosely, the model assigns a larger probability of
crossover (speaking very loosely again), as the
error term is more likely to be larger, as
represented in the picture by a longer dashed line segment.

Consider our earlier framework of thinking of each summand of the log
likelihood function as giving a penalty for crossover (and contributing
a small negative number to the sum if not). Recall from
(\ref{loglikelihood}) that this penalty is always negative. Suppose that there is crossover for observations
$1$ and $2$; from the graph, we can see that these two points are in the
region $\{ \bx : \bx' \bbeta < 0 \}$, so in other words, given that we
are assuming there is crossover, we have $y_1=y_2=1.$\footnote{As with the probit graph, one needs
to know the value of $\beta$ to determine which side of the line is the
region $\{ \bx : \bx' \bbeta < 0 \}$ and which side the region $\{ \bx :
\bx' \bbeta >0 \}.$ Here we have in fact chosen
$\bbeta=(-1.2,1)$. Obviously had we chosen $(1.2,-1)$, or some positive
scalar multiple of it, the regions would be reversed.} From
(\ref{loglikelihood}), the penalty for crossover, as $y_1=y_2=1$, is 
\[ \ln \left( \Phi \left( \frac{ \bx_i' \bbeta}{\exp(\bz_i' \bgamma)}
  \right) \right) \qfor i=1,2,
\]
which is, speaking very loosely, smaller (in magnitude), the larger is $\exp(\bz_i' \bgamma)$ (as $\bx_i'
\bbeta<0$ for $i=1,2.$). Of course, the size of the penalty also depends
on the magnitude of $\bx_i' \bbeta$, and more precisely on the ratio
$\dspl{\frac{ \bx_i' \bbeta}{\exp(\bz_i' \bgamma)}}$. It is not
immediately clear from the graph, but for our purpose here of providing
exposition, one can perhaps see that it makes some intuitive sense that
this penalty would be smaller (in magnitude) for observation $1$ than for observation
$2$.\footnote{To see this precisely, first recall that we have $\bx_1'
      \bbeta <0$ and $\bx_2' \bbeta <0$. The fact that the dashed
  line for observation $1$ crosses the line $\bx' \bbeta=0$ shows, from (\ref{vdefs}), that
  $\exp(\bz_1' \bgamma)> |\bx_1' \bbeta|$, whence $\dspl{\sol{\bx_1'
      \bbeta}{\exp(\bz_1' \bgamma)} }>-1$ holds. On the other hand, the dashed
  line for observation $2$ does not cross the line $\bx' \bbeta=0$. Thus
  $\exp(\bz_2' \bgamma)< |\bx_2' \bbeta|$, whence $\dspl{\sol{\bx_2'
      \bbeta}{\exp(\bz_2' \bgamma)} }<-1$ holds. Therefore we have
  $\dspl{\sol{ \bx_2' \bbeta}{\exp(\bz_2' \bgamma)} <\sol{ \bx_1'
      \bbeta}{\exp(\bz_1' \bgamma)}}$ and thus
  $\dspl{\ln \left( \Phi \left( \sol{ \bx_2' \bbeta}{\exp(\bz_2' \bgamma)}
  \right) \right) < \ln \left( \Phi \left( \sol{ \bx_1' \bbeta}{\exp(\bz_1' \bgamma)}
  \right) \right)}$. So the
penalty for observation $2$ is more negative than that for observation $1$, i.e., the penalty for observation $1$ is
smaller (in magnitude) than the penalty for observation $2$.

We present an example showing why the perpendicular distance from the end of
the dashed line to the line $\bx' \bbeta=0$ does not completely determine the penalty,
although it may seem at first that this might be the case. Consider two
observations indexed by $a$ and
$b$ with $\bx_a' \bbeta=5$ and $\exp(\bz_a' \bgamma)=1$, and $\bx_b'
\bbeta=25$ and $\exp(\bz_b' \bgamma)=15$. For each of these two observations, as
$\bx_i' \bbeta>0$ holds for $i=a,b$, the endpoint of the dashed line
that is closest to the line $\bx' \bbeta=0$ is $v_{i_1}$ in the notation
of (\ref{vdefs}), and thus the distance from the end of each
dashed lines to the line $\bx' \bbeta=0$ is determined by $\bx_i'
\bbeta- \exp(\bz_i' \bgamma)$. Set $u_i=\bx_i' \bbeta- \exp(\bz_i'
\bgamma)$. This distance is a strictly increasing
function in $u_i$ with value $d(u_i)$; i.e. it
varies in an increasing fashion (albeit nonlinearly) with $u_i$. Now for observation $a$, this distance is
$d(4)=d(5-1)$, whereas it is $d(10)=d(25-15)$ for observation $b$. The
latter is larger, so one might be tempted given the way we have
represented the model graphically to believe that the penalty
would be smaller for observation $b$, as the dashed line is farther from
the line $\bx' \bbeta=0$. But in fact, the ratio $\dspl{\sol{ \bx_i'
    \bbeta}{\exp(\bz_i' \bgamma)}}$ is larger for $i=b$ than it
is for $i=1$, as $\dspl{\sol{15}{25}>\sol{1}{5}}$, so the penalty for
observation $b$ is in fact larger. We could have drawn the dashed lines
with length equal to the above ratio, but despite the flaws, we decided
to draw the dashed lines as here, as it gives a representation
of the range of values for the latent variable $\dspl{y^{\ast}}$, and
because other alternatives seemed less natural and less apt to provide
the reader with graphical
intuition.} Thus a maximizer gets
penalized much less for choosing $\bbeta$ so that $\bx_1$ is on the
wrong side of the dividing line than for choosing $\bbeta$ so that
$\bx_2$ is on the wrong side.

\begin{figure*}[!h]
\centering
\includegraphics[scale=0.5]{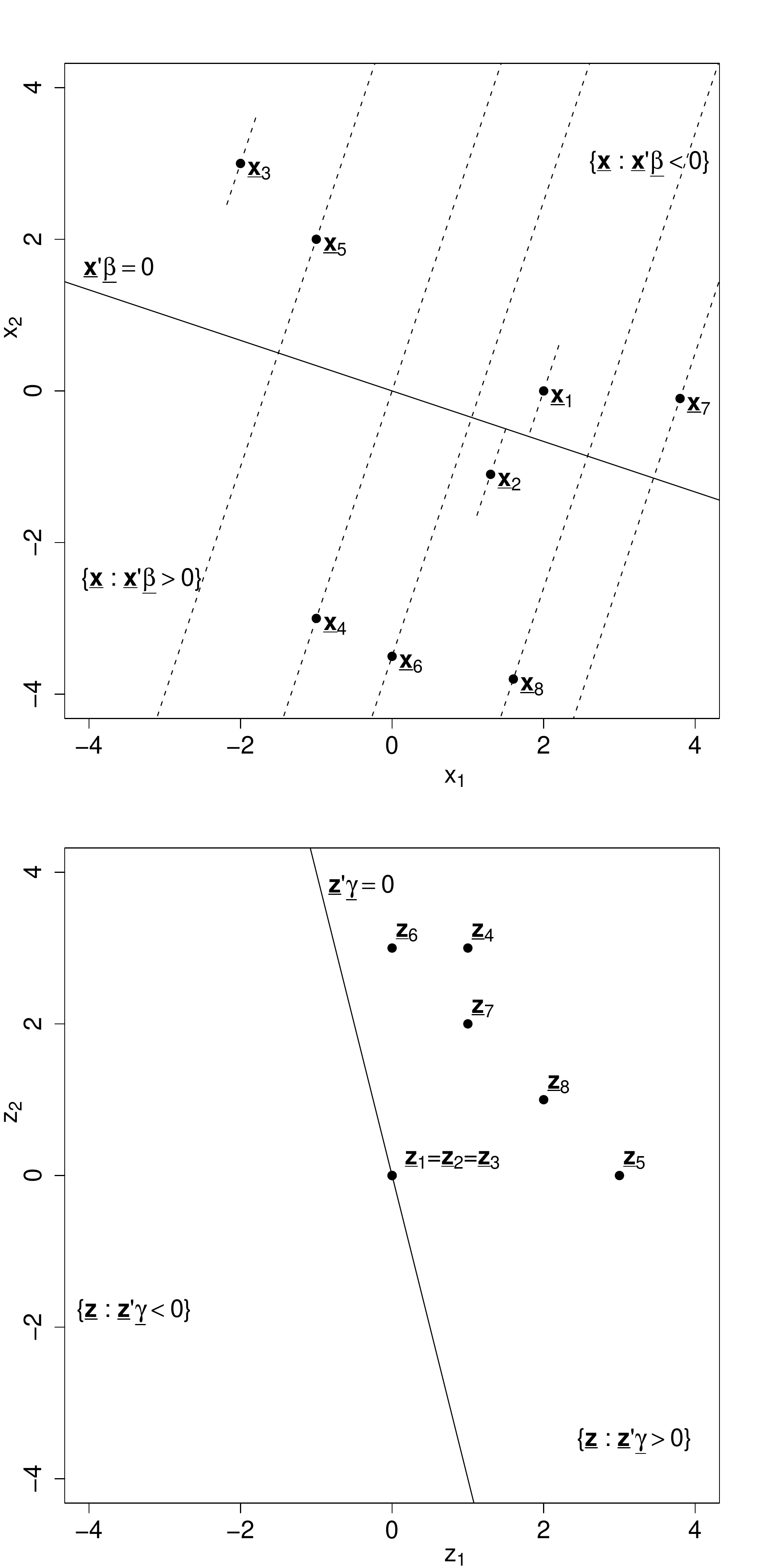}
\caption{A Graphical Representation of a Plateau Solution in the
  Heteroskedastic Probit Model}
\label{fig:plateauinterpret}
\end{figure*}

Now consider the bottom graph in Figure \ref{fig:hetprobitinterpret},
which depicts the observations $\bz_i \ (1 \leq i \leq 5)$. Also we see
the hyperplane $\bz' \bgamma=0$ (here a line, as we have assumed above
that $\bgamma$ is two-dimensional). Here the upper right half represents
the region $\{ \bz : \bz' \bgamma>0 \}$, and the lower left half plane
the region $\{ \bz : \bz' \bgamma<0 \}$.\footnote{Here we have chosen
  $\bgamma=(1,5)$.} Note that there is no special
significance of the line $\bz' \bgamma=0$ analogous to the significance of the line
$\bx' \bbeta = 0$, but we include it as an indicator of the value of $\bz' \bgamma$. Observe that those observations $i$ with $\bz_i' \bgamma$ positive
and large will have $\exp(\bz_i' \bgamma)$ large and thus have a longer
associated dashed line segment in the top graph of Figure
\ref{fig:hetprobitinterpret}. 

An optimizer, then, attempts to simultaneously choose two lines
(or, more generally, two hyperplanes): one to split the $\bx_i$'s into
two regions to try to give observations a positive value for $\bx_i'
\bbeta$ if $y_i=1$ and a negative value if $y_i = 0$, and the second to
arrange the $\bz_i$'s so that (loosely speaking) the smaller is $\bz_i'
\bgamma$, the less likely observation $i$ is to exhibit crossover. 

Now observe what happens for an optimizer who chooses a plateau
solution. In this case, we return to the earlier assumption that $Z \geq 0$ (which is of course not the case in the bottom
graph in Figure \ref{fig:hetprobitinterpret}), and that there are some
observations $i$ with $\bz_i = \bzero$. For this plateau solution, the optimizer
chooses $\bgamma$ with large and positive components.\footnote{The
  choices we have made for these graphs, for ease of illustration, are $\bbeta=(-1.5,-5)$ and $\bgamma=(8,2)$.} The two planes for
an illustrative case are
presented in Figure \ref{fig:plateauinterpret}. First look at the bottom
panel. Observations with $\bz_i=\bzero$, which we label 
$\bz_1,  \bz_2$ and $\bz_3$ for expositional convenience, of course lie
directly on the line, while all of the other observations (represented
here by unlabeled points) lie in the region $\bz_i' \bgamma > 0$  and
away from the line. In the top graph, we see that only the
points $\bx_1, \bx_2$ and $\bx_3$ have small associated dashed line
segments, while all others have large associated segments. (Note that
some of these segments extend outside of the region of the graph.)
So, by virtue of choosing $\bgamma$ with large, positive components, the
optimizer has ensured that only the observations $i=1,2,3$ matter very
much for choosing the line $\bx' \bbeta = 0$, i.e. for choosing $\bbeta$.

\section{Potential Improvement and Other Suggestions}
\label{sec:suggestion}

We now propose a straightforward method to mitigate the problems we have
identified in the heteroskedastic probit model. We suggest ensuring that
the model simply satisfies the property that at least one of the
components of the vectors in $\bz_i$ take both signs. We would hope for this to avoid the plateau
problem, or at least the exact form of it described above. Observe, however, that we do not claim that if one
merely does so then maximizing the log likelihood function for the heteroskedastic probit model is
robust; we are only stating that this is likely a helpful modification. Here we assume that there is no constant term in $\bz$.\footnote{There
  may or may not be a constant term in $\bx$ here. In general, there should of course
  not be a constant term in {\it both} $\bx$ and $\bz$, as the model
  would not be identified.} If there
were a constant term in $\bz_i$, say $z_{i1}$, one could simply choose the
parameter $\gamma_1$ to be positive and very large relative to the other
components of $\bz_i$ and thus ensure that $\exp(\bz_i' \bgamma)$ is always
large, thus undoubtedly leading to a similar plateau problem.

We give a demonstration of the potential efficacy of our solution, by comparing the performance of an
optimization algorithm on (i) a data set with $Z \geq 0$ and (ii)
a translation of this first data set that ensures that the observations $\bz_i$ take both
signs. The first data set is created as described above, and, for
example, is forced to satisfy the requirement that the percentage of observations with
crossover is between 20\% and 30\%. Denote the data set by
$\mathcal{D}$. Let $n$ be the number of observations in the data set. $\mathcal{D}$
consists of the parameters $\bbeta_0$ and $\bgamma_0$ and the
observations $\by$, $X$ and $Z$, so we write
\[ \mathcal{D} = \{ \bbeta_0, \bgamma_0, \by, X, Z \}.
\]
Also define $k_1=\dim(\bbeta_0)$ and $k_2=\dim(\bgamma_0)$. 
Observe that we have $\dim(\by)=n$, while $\dim(X)=n \times k_1$ and $\dim(Z)=n
\times k_2$. 
Write 
\[ \bgamma_0 = (\gamma_{01}, \gamma_{02}, \ldots,
\gamma_{0k_2}).
\] 
For convenience, write
\[ \mathbb1 = \underbrace{(1,1, \ldots, 1)}_{\mbox{\scriptsize{length} }
  k_2}.
\]
Now we define the transformed data set
$\widetilde{\mathcal{D}}$. Let
\begin{eqnarray*} \widetilde{\bbeta}_0 & = & \left( \exp \left(
      - \sum_{j=1}^{k_2} \frac{\gamma_{0j}}{2} \right) \right) \bbeta_0, \\
  \widetilde{\bgamma_0} & = & \bgamma_0 ,\\
  \widetilde{\by} & = & \by, \\
  \widetilde{\bx}_i & = & \bx_i \qfor 1 \leq i \leq n, \\
  \widetilde{\bz}_i & = & \bz_i - \frac{1}{2}
  {\mathbb{1}} = \bz_i - \frac{1}{2} \underbrace{(1,1, \ldots,
    1)}_{\mbox{\scriptsize{length} } k_2} \qfor 1 \leq i \leq n, \\
\end{eqnarray*}
and 
\[ \widetilde{\mathcal{D}} = \left\{ \widetilde{\bbeta}_0, \widetilde{\bgamma}_0,
\widetilde{\by}, \widetilde{\bx}, \widetilde{\bz}  \right\}.
\]
We note for emphasis that we have
$\dim\left(\widetilde{\bbeta}_0\right)=k_1$ and
$\dim\left(\widetilde{\bgamma}_0\right)=k_2,$ and also $\dim\left(\widetilde{\by}\right)=n$, while $\dim\left(\widetilde{X}\right)=n \times k_1$ and $\dim\left(\widetilde{Z}\right)=n
\times k_2$. 

Now, consider, for each $i$, the expressions $\dspl{\frac{ \bx_i' \bbeta}{\exp(\bz_i'
    \bgamma)}}$ and $\dspl{\frac{ \widetilde{\bx}_i' \widetilde{\bbeta}}{\exp(\widetilde{\bz}_i'
    \widetilde{\bgamma})}}$. These appear as terms in each summand of
the respective log likelihood functions for the heteroskedastic probit model, for the
original and transformed data sets. We claim that for all $i$, the value
of this term for the original data set equals its value for the
transformed data set at each data set's respective model parameters: for
we have
\begin{eqnarray*} \frac{ \widetilde{\bx}_i' \widetilde{\bbeta_0}}{\exp(\widetilde{\bz}_i'
    \widetilde{\bgamma_0})} & = & \left( \exp \left( -
      \sum_{j=1}^{k_2} \frac{\gamma_{0j}}{2} \right) \right) \frac{
    \bx_i' \bbeta_0}{\exp\left( \left(\bz_i
        - \frac{1}{2} \mathbb1 \right)'  \bgamma_0 \right)} \\
  & = & \left( \exp \left( -
      \sum_{j=1}^{k_2} \frac{\gamma_{0j}}{2} \right) \right) \frac{
    \bx_i' \bbeta_0}{\exp \left(\bz_i' \bgamma_0 \right) \exp \left(
        - \frac{1}{2} \mathbb1 '  \bgamma_0 \right)} \\
  & = & \frac{ \bx_i' \bbeta_0}{\exp \left(\bz_i' \bgamma_0 \right)}. \\
\end{eqnarray*}
Together with the observation that $\widetilde{\by}= \by$, we can see
that the $i^\ith$ summand of the log likelihood function for the
original data set at its model parameters $(\bbeta_0, \bgamma_0)$ is
equal to the $i^\ith$ summand of the log likelihood function for the
transformed data set at its model parameters $(\widetilde{\bbeta}_0,
\widetilde{\bgamma}_0)$. In other words, writing $\dspl{\by=
  \left(y_1, y_2, \ldots, y_n \right)}$ and $\dspl{\widetilde{\by}=
  \left(\tilde{y}_1, \tilde{y}_2, \ldots, \tilde{y}_n \right)}$, we have:
\begin{align*} \left( y_i \ln \Phi
  \left( \frac{\bx_i' \bbeta}{\exp( \bz_i'\bgamma )} \right) + (1- y_i)
  \ln \left( 1 - \Phi
  \left( \frac{\bx_i' \bbeta}{\exp( \bz_i'\bgamma )} \right) \right)
\right) & = \\
&  \hskip -150pt \left( \tilde{y}_i \ln \Phi
  \left( \frac{\widetilde{\bx}_i' \widetilde{\bbeta_0}}{\exp( \widetilde{\bz}_i'\widetilde{\bgamma}_0 )} \right) + (1- \tilde{y}_i)
  \ln \left( 1 - \Phi
  \left( \frac{\widetilde{\bx}_i' \widetilde{\bbeta}_0}{\exp( \bz_i'\widetilde{\bgamma}_0 )} \right) \right)
\right) \\
& \hskip -50 pt \qfor 1 \leq i \leq n. \\
\end{align*}
It of course follows immediately that the respective log likelihood
functions at the respective model parameters are equal, i.e. that
\[ \ell \left(\bbeta_0, \bgamma_0 \mid y, X, Z \right)= \ell
\left(\widetilde{\bbeta}_0, \widetilde{\bgamma}_0 \mid \widetilde{y},
  \widetilde{X}, \widetilde{Z} \right). 
\]

Now we turn to some details. As described above, we create a pseudorandom set $\mathcal{D}$. We then
generate 1000 random choices of initial values in the box
$\dspl{[-5,5]^5}$, and measure the search performance of the BFGS algorithm in estimating the model parameters
$\dspl{\left(\bbeta_0, \bgamma_0 \right)}$. Then we
transform the data set $\mathcal{D}$ as described above to get a new
data set $\dspl{\widetilde{\mathcal{D}}}$. We then transform these 1000 new
random choices of initial values (in the same way we transformed
$\bbeta$ in the equations above, keeping $\bgamma$ unchanged), and
subsequently measure the search performance of the BFGS algorithm in approaching its model parameters
$\dspl{\left(\widetilde{\bbeta}_0, \widetilde{\bgamma}_0
  \right)}$.\footnote{We also
  ran a simulation on the same data set but where we chose different
  initial values for each of the original and transformed data set. The
  results are very similar.}

\begin{figure*}[!h]
\centering
\includegraphics[width=0.98\textwidth]{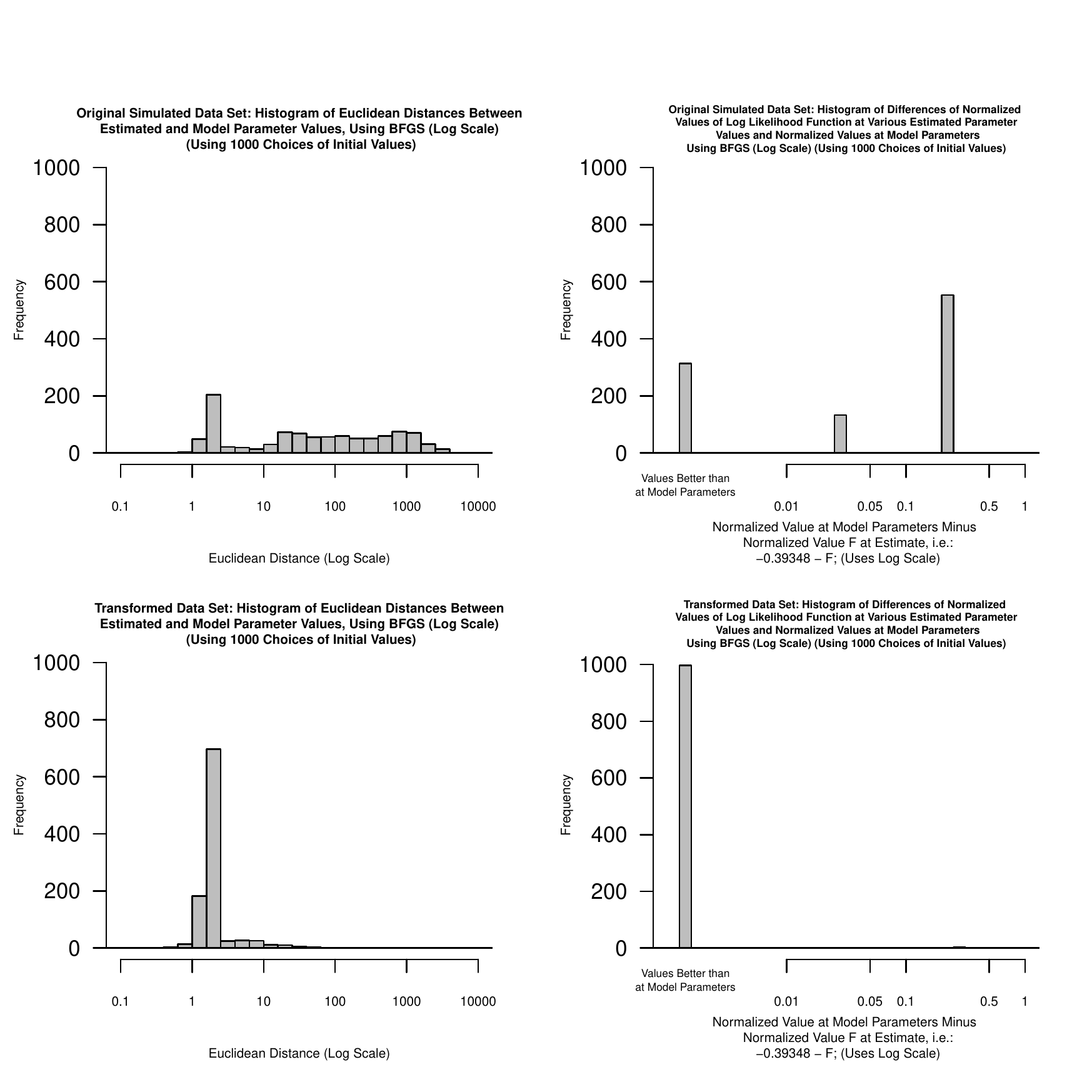}
\caption{Comparison of Performance of the BFGS Search Algorithm for a
  Simulated Data Set and a Transformation of that Data Set (Using the
  Heteroskedastic Probit Model), for 1000 Random Choices of Initial Values} 
\label{fig:transformationcomparison}
\end{figure*}

We present the results in Figure
\ref{fig:transformationcomparison}. First consider the left
graphs. The top left plot displays a histogram of the Euclidean
distances between the estimated and model parameter values for the
$1000$ runs for the original data set $\mathcal{D}$; the bottom left
plot displays the analogous histogram for the transformed data set
$\dspl{\widetilde{\mathcal{D}}}$.\footnote{Observe that the plots have
  the same scales on the horizontal and vertical axes.} Clearly the
performance in the latter case is far superior. Now consider the right
column of graphs. As with earlier plots, we also display histograms of the
(signed) differences between the normalized values of the log likelihood at the
model and estimated parameters. (Recall that the former might generally
be expected to be greater, i.e. a smaller negative number.) The top
graph corresponds to the original simulated data set; the bottom plot to
the transformed data set. Horizontal and vertical axes in both
graphs have the same scale. Again, in both graphs, we have also
plotted a bar that represents the number of choices of initial values for
which the log likelihood value at the resulting estimates is actually
larger than the log likelihood at the model parameters.\footnote{As
  above, the horizontal position of this bar is of course not
  meaningful.} There are many more of these for the transformed data
set, and in fact for the transformed data set, there are only three choices
of initial values (out of $1000$) for which the log likelihood at the
estimated parameters is worse than (i.e. less than) the log likelihood
at the model parameters. Clearly, by this method of comparison as well,
the BFGS search algorithm performs much better on the transformed data set. 

Naturally, we have only presented evidence for one simulated data set
and one search algorithm.\footnote{We also performed similar simulations
  for the CG, Nelder-Mead and SANN algorithms. The CG algorithm performs
much better for the transformed data set than for the original
data set, and the Nelder-Mead
method performs better but not strongly so; the SANN algorithm on the other hand performs fairly
well for both data sets.} But we have given reasons why this
transformation may mitigate the plateau problem.\footnote{It may be
    of interest to discuss, in terms of a search algorithm, how the
    transformation affects  
    the plateau problem. Note that parameter values on the plateau for the
    original set, i.e. with large values of $\bgamma$, correspond, under
    the transformation, to values of $\bbeta$ close to $\bzero$; this can
    be seen from the definition of the transformation. Thus, small jumps
    in a local search on the transformed set would lead a search
    algorithm away from $\bzero$, which would correspond to escaping the
    plateau in the parameter space for the original data set.}   


\subsection{Additional Suggestions}

In addition to the above modification, we make a few other suggestions
that may be helpful for researchers to bear in mind when using a search
algorithm to estimate a heteroskedastic probit model. 

One thing we suggest is that a researcher should try multiple different
starting values for search algorithms. One could then check if one gets
a stable set of resulting parameter estimates; if not, caution is called
for. This is presumably a good approach in general, but it appears to be especially warranted for the
heteroskedastic probit model. As well, if a researcher uses a search
algorithm, and finds an estimate $\widehat{\bgamma}$ that has large
positive components, there should be concern that it is a plateau
solution; in line with the previous suggestion, the researcher should
try many other initial values. If that does not improve the situation,
results should be interpreted with caution.  The simulated annealing algorithm (SANN in this
implementation) is, in some respects, similar in
spirit to the notion of starting search algorithms at multiple initial
values, in that it incorporates random jumps in the procedure. In some
of the simulations SANN performed much better than some of the
other algorithms; however, it did not perform as well as other methods
on the Alvarez-Brehm data. So a simulated annealing algorithm might be a
useful search method for the heteroskedastic probit. We emphasize that we do not
view a simulated annealing algorithm as a panacea by any means, and we are not implying
that simply using a simulated annealing method suffices to address the problems we
have identified in the heteroskedastic probit model.

Finally, we make an observation regarding the special case in which
$\bx$ and $\bz$ are independent. From (\ref{plateauform}),
we can see that intuitively, for positive and large choices of
$\widehat{\bgamma}$, it seems that the choice of $\widehat{\bbeta}$ would be determined by
the optimal choice of $\widehat{\bbeta}$ for the subset of $i$ for which $\bz_i =
\bzero$. If $\bx$ and $\bz$ are independent, then the observations $\bx_i$
for $i$ in this subset are a random sample, and thus we would expect the
estimate of $\bbeta_0$ to be good in some sense if the estimate $\widehat{\bgamma}$ is
positive and large. Thus this presents a potential (informal) method of estimation
for the heteroskedastic probit in this special case: first, find an estimate
$\widehat{\bbeta}$ and if the associated estimate $\widehat{\bgamma}$
is positive and large, then, second, fix this value of $\widehat{\bbeta}$ and
use a search algorithm to estimate only the $\bgamma_0$ parameter.\footnote{An
  alternative method might involve starting with a positive and large
  initial value for $\bgamma$. Also, we have observed the estimate of $\bbeta_0$
  to be close to the model or A-B parameters in multiple
  simulations. But we do not investigate making either of the proposed
  methods more formal, as this is not the focus of this paper.}



\section{Final Remarks}

Here we have explored how commonly used optimization algorithms perform
when applied to the heteroskedastic probit likelihood. Using both
simulated data sets and in a re-analysis of the seminal work by
Alvarez and Brehm, we find that some optimization algorithms can often converge at
incorrect solutions. We compare these same algorithms when applied to
standard probit log likelihood functions and find no such problems.
We also sketch a heuristic argument for why these difficulties may be inherent in the heteroskedastic
probit likelihood. Our work constitutes a first step, by calling
  attention to some potential problems in using the model. But more remains to be done to understand the
  particular functional form. However, analysts that estimate heteroskedastic probit
models should at the very least ensure that a variety of starting values
converge to the same set of estimates. They may also want to perform
  a transformation of the data and parameters in the fashion we have
  described. It would also be advisable to try more robust optimization algorithms such as simulated annealing or a genetic optimization algorithm \citep{Mebane:1998}.

\clearpage
\bibliographystyle{myapsr}
\bibliography{keele_revised2}

\clearpage

\appendix
\section{Appendix: Graphs for the SANN Search Algorithm}

\begin{figure*}[!h]
\caption{Performance of the SANN Search Algorithm on a
    Simulated Data Set for the Heteroskedastic Probit Model, for 1000 Random Choices of Initial Values}
\centering
\includegraphics[width=0.98\textwidth]{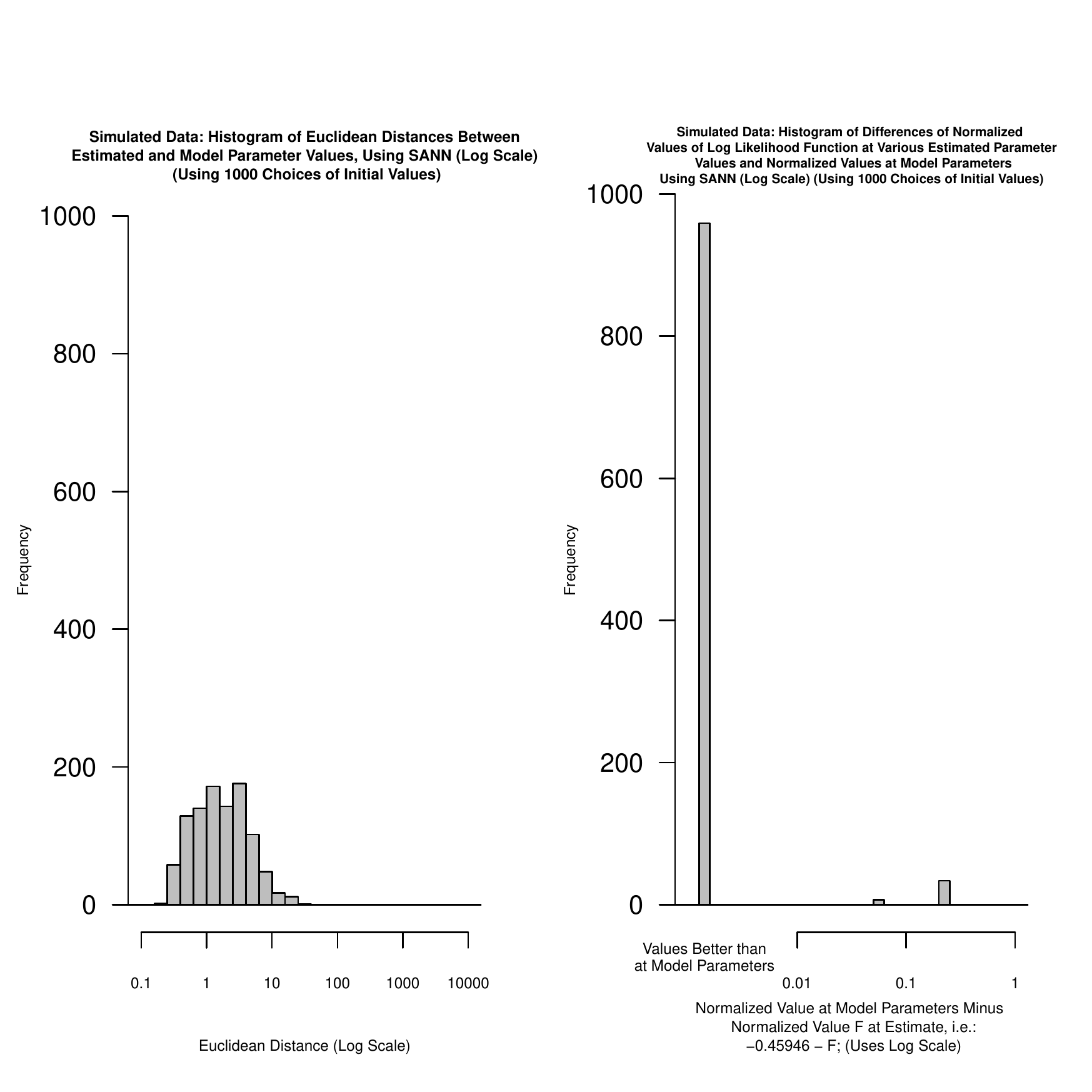}
\label{fig:simdata3}
\end{figure*}

\newpage
\clearpage

\begin{figure*}[!h]
\caption{Performance of the SANN Search Algorithm on a Simulated Data
  Set for the Probit Model, for 1000 Random Choices of Initial Values}
\centering
\includegraphics[width=0.98\textwidth]{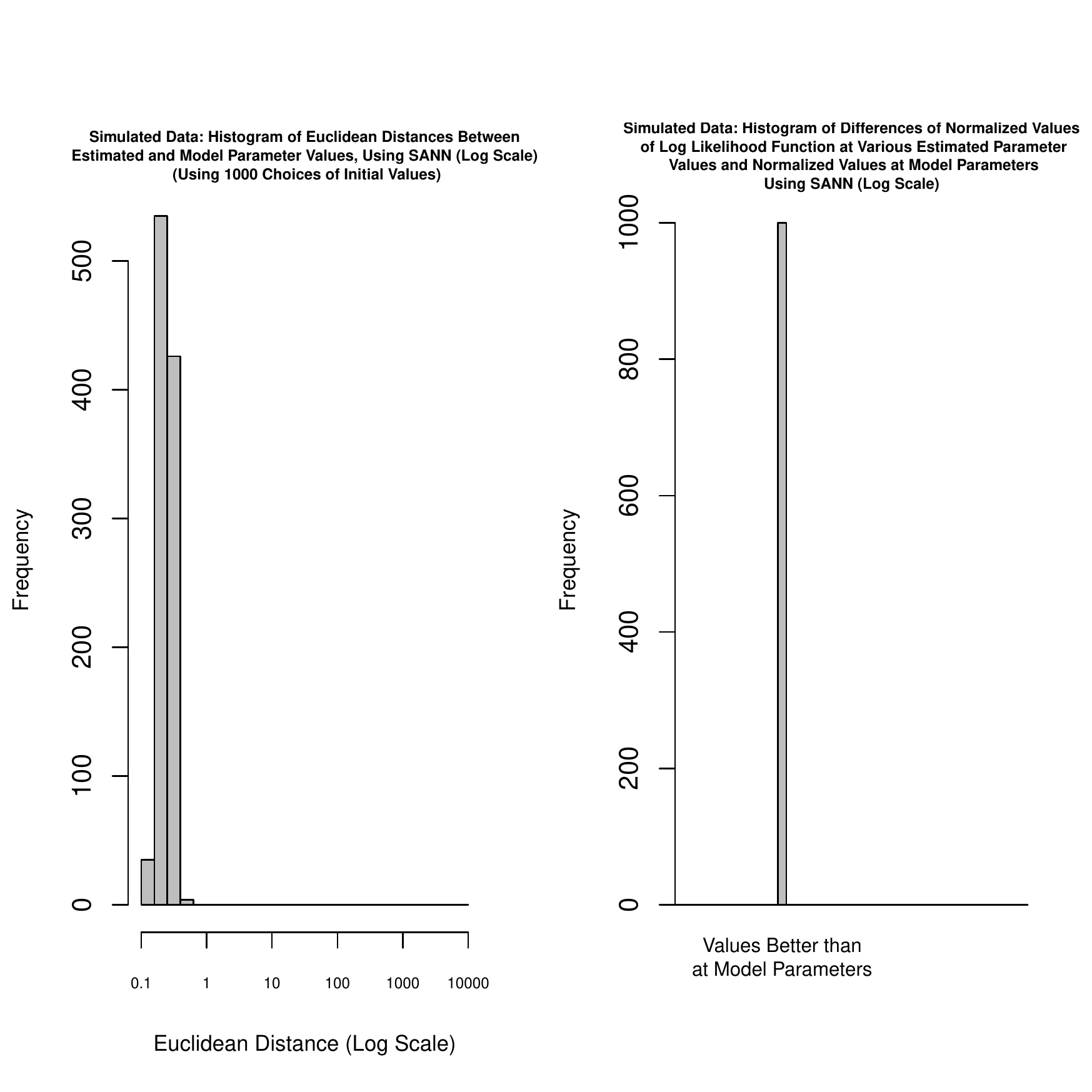}
\label{fig:simdataprobit3}
\end{figure*}

\newpage
\clearpage

\begin{figure*}[!h]
\caption{Performance of the SANN Search Algorithm on the
    Alvarez-Brehm Data, for 1000 Random Choices of Initial Values}
\centering
\includegraphics[width=0.98\textwidth]{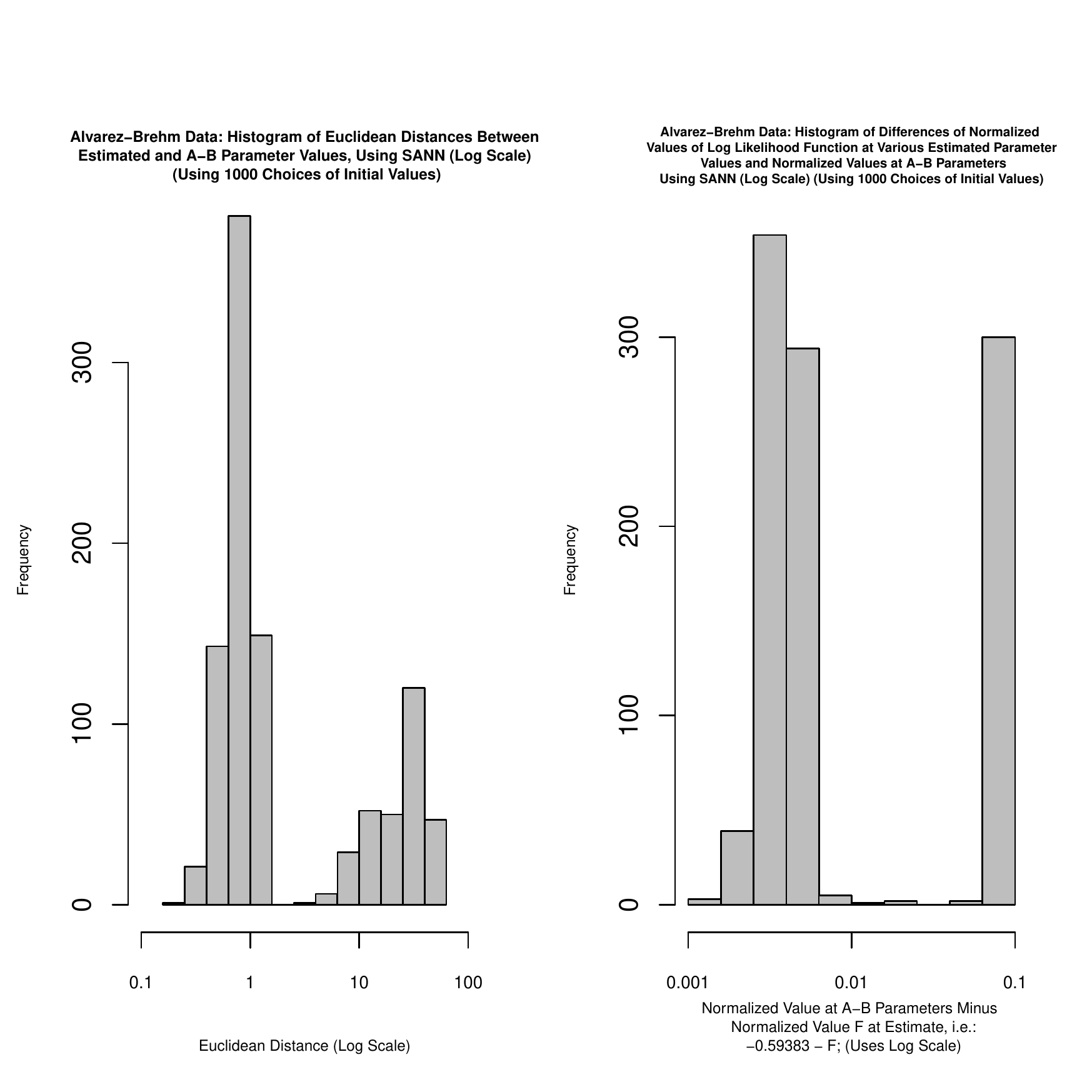}
\label{fig:abdata1}
\end{figure*}





\end{document}